\documentclass[aps, prb, preprint, amsmath, amssymb]{revtex4}
\usepackage{graphicx, subfigure, times, setspace}
\begin{document}

\title{Atomistic simulations of rare events using gentlest ascent dynamics}

\author{Amit Samanta}
\email{asamanta@math.princeton.edu}
\affiliation{Program in Applied and Computational Mathematics,
  Princeton University, Princeton, New Jersey 08544, USA}

\author{Weinan E}
\email{weinan@math.princeton.edu}
\affiliation{Department of Mathematics and Program in Applied and
  Computational Mathematics, Princeton University, Princeton, New Jersey 08544, USA}

\date{\today}

\begin{abstract}
The dynamics of complex systems often involve thermally activated
barrier crossing events that allow these systems to move from one
basin of attraction on the high dimensional energy surface to
another. Such events are ubiquitous, but challenging to simulate using
conventional simulation tools, such as molecular dynamics. Recently,
Weinan E et al. [Nonlinearity, $\bf 24$(6),1831(2011)] proposed a set
of dynamic equations, the gentlest ascent dynamics (GAD), to describe
the escape of a system from a basin of attraction and proved that
solutions of GAD converge to index-1 saddle points of the underlying
energy. In this paper, we extend GAD to enable finite temperature
simulations in which the system hops between different saddle points
on the energy surface. An effective strategy to use GAD to sample an
ensemble of low barrier saddle points located in the vicinity of a
locally stable configuration on the high dimensional energy surface is
proposed. The utility of the method is demonstrated by studying the
low barrier saddle points associated with point defect activity on a
surface. This is done for two representative systems, namely, (a) a
surface vacancy and ad-atom pair and (b) a heptamer island on the
$(111)$ surface of copper. 
\end{abstract}

\maketitle

\section{Motivation}

The dynamics of many complex systems proceed via sequences of
infrequent transition events from one metastable state to
another. Well-known examples include chemical reactions, conformation
changes of bio-molecules, nucleation events during phase transition,
etc.$\cite{Eyring35, HanggiTB90, Fecht92}$ Metastability is
characterized by the appearance of disparate time scales. Two most
important time scales are the relaxation time scale of a state, and
the transition time out of the state. A state is metastable if the
second time scale is much bigger than the first. For this reason,
transitions between different metastable states are rare events. If
the system has a rather simple potential energy landscape (PES), the
bottleneck for the transition is a saddle point (generally of
index-1), called the transition state, of the potential energy of the
system.  

Crippen and Scheraga pioneered numerical algorithms for climbing out
of basins of attraction.$\cite{CrippenS71}$ In their method, starting
from an initial point ${\bf x}_{0}$, in the $k$-th step, the system
translates along the direction ${\bf r} = \left({\bf x}_{k} - {\bf
    x}_{0}\right)$ by a suitable step length $\epsilon$ to yield ${\bf
  x}_{k}^{\ast} = {\bf x}_{k} + \epsilon{\bf r}$. This is followed by
minimization on a hyperplane $S_{k}$ perpendicular to $\bf r$ to yield
${\bf x}_{k+1} = \underset{{\bf x}\in
  S_{k}}{\operatorname{argmin}}\;\;V\left({\bf
    x}_{k}^{\ast}\right)$. This process is repeated until the system
reaches a saddle point. To sample different saddle points, the authors suggested
using the eigenvectors $\{{\bf w}_{1}, {\bf w}_{2}, {\bf w}_{3},..., {\bf w}_{N}\}$
of the Hessian at ${\bf x}_{0}$, i.e. ${\bf x}_{0}^{\ast} = {\bf
  x}_{0} + \epsilon{\bf w}_{i}$, $i\in\left[1, N\right]$, followed by
minimization on $S_{0}$ to obtain ${\bf x}_{1}$. 

Later works on exploring the high dimensional potential energy surface
(PES) are based on the realization that evaluating minimum eigenmode of the
Hessian is central to the convergence to an index-1 saddle
point. The system can however, start from the initial locally stable
point by following any eigenmode of the Hessian or other randomly
selected direction vectors.$\cite{CrippenS71, BarkemaM96, PedersenHJ11}$ For
smaller systems, an eigenvalue problem of the Hessian is easy to
solve by diagonalization (though one has to remove the global rotational and
translational components), but this becomes computationally
challenging for systems of higher dimensions. Consequently, many
techniques have been proposed to evaluate the minimum eigenmode of the
Hessian in an optimal fashion. 

For instance, Cerjan and Miller suggested a technique that requires
selecting a trust region around a point on the multidimensional PES
and approximating the energy of the system within this trust region by
a quadratic expression.$\cite{CerjanM81}$ An optimal direction to
translate the system is then determined by evaluating the extremum of the
energy on the boundary of the trust region. The key to proper
evaluation of the optimal direction lies on proper selection of the
trust region - if the trust region too small, then the linear term
dominates and only one minimum is obtained; on the other hand, for a large
trust region the quadratic approximation of the PES becomes poor. The
computational effort, for this method, scales up rapidly as more
degrees of freedom are included in the system and at the same time the
construction of the Hessian can become a formidable computation
challenge when analytical second derivatives are not available.  

A more economic approach to evaluate the ascent direction, that avoids
the constructing the full Hessian, is the dimer method proposed by
Henkelman and J$\rm\ddot{o}$nsson.$\cite{HenkelmanJ99}$ A dimer
consists of two images of the configuration separated by a small
distance ($\sim 10^{-2}$ $\rm\AA$). The dimer translates towards a
saddle point by the modified force $\bf F_{R} - 2\bf F^{\parallel}$,
where $\bf F_{R}$ is the net force acting on the dimer and $\bf
F^{\parallel}$ is the component of the force parallel to a dimer
direction $\bf n$. The direction $\bf n$ is determined by minimizing
the dimer energy, i.e. the direction vector is updated after $k$-th
iteration by
\begin{equation}
  {\bf n}_{k+1} = \underset{{\bf n},\;\;{\left|\bf
        n\right|^{2}=1}}{\operatorname{argmin}}\;\;\left[V\left({\bf
        x}_{k}+\epsilon\bf n\right) + V\left({\bf x}_{k}-\epsilon\bf n\right)\right]
\end{equation}
where, $V\left(\bf x\right)$ is defines the PES. In the above,
$V\left({\bf x}_{k}+\epsilon\bf n\right)$ and $V\left({\bf
    x}_{k}-\epsilon\bf n\right)$ represents the energy of the dimer
separated by a distance of $2\epsilon$. 
The process of obtaining the minimum eigenmode of the Hessian is
implemented by first selecting a suitable plane spanned by the 
rotational force acting on the dimer and the direction $\bf n$. The
dimer is then rotated in this plane to obtain an optimal $\bf n$
corresponding to a minimum energy of the dimer.$\cite{OlsenKHAJ04}$
A dynamical version of the dimer method was recently proposed by
Poddey et al.$\cite{PoddeyB08}$

Similarly, Munro and Wales suggested an iterative scheme to calculate
the minimum eigen mode of the Hessian.$\cite{MunroW99}$ Starting from
an initial guess direction ${\bf n}_{0}$, the eigenvector
corresponding to the minimum eigenvalue of the Hessian is found
approximately by successive operation of the shifted Hessian
i.e. ${\bf n}_{k} = \left({\bf H} - \lambda\mathbb{I}\right)^{k}{\bf
  n}_{0}$, where, $\lambda$ is a constant and $\mathbb{I}$ is the
identity matrix. In this procedure, since we only need the product of
the Hessian with a vector, there is no need to explicitly calculate
the Hessian thus reducing the computational complexity
significantly. Once the minimum eigenmode is determined the system is
then translated in the configuration space by a suitable step length.  

Other PES exploration algorithms like the activation-relaxation
method (ART) proposed by Barkema and Mousseau$\cite{BarkemaM96}$ do
not guarantee visiting an index-1 saddle point. Analysis of the 
convergence of ART to saddle points is detailed in the Appendix. Even
though ART is an excellent exploration tool, the saddle points sampled
may not correspond to the low barrier saddle points that drive rare
transition events. Proposals for fixing this problem can be found in
recent literature.$\cite{MousseauB00, CancesLMMW09, MarinicaWM11}$
 
The philosophy that underlies the gentlest ascent dynamics (GAD) is to
take one step back and formulate a set of dynamic equations whose
solutions converge to saddle points. Numerical algorithms can then be
constructed by discretizing this set of dynamic equations. As we
illustrate in this paper, having a set of dynamical equations helps us
to analyze the stability and convergence close to different fixed
points on the PES. At the same time, it is easy to extend such a
formalism to finite temperature cases. In this paper, we will discuss
how GAD can be naturally extended as a tool for sampling saddle points
and for the exploration of the configuration space. In particular, we
will propose finite temperature molecular dynamics version of GAD,
MD-GAD.  MD-GAD can be used to produce trajectories that hop from
saddle point regions to saddle point regions.  Such time series can be
useful for a variety of purposes. We also discuss practical issues
related to GAD such as selecting initial conditions, etc. for high
dimensional systems. 

The arrangement of the paper is as follows : in
Section II, we discuss the set of dynamical equations corresponding to
finite temperature and zero temperature conditions to sample saddle
points. In Section III, we discuss the different ways to implement the
set of dynamical equations, namely, identifying an important region
associated with a rare transition event, proper sampling of direction
to move out of the energy well and selection of convergence
criteria. Next, we use two examples -$(i)$ a heptamer island on
$(111)$ surface of copper and $(ii)$ a surface vacancy and ad-atom
pair on the $(111)$ surface of copper to study the relevant saddle
configurations and the energy barriers. 

\section{Equations of motion}

We assume that the system being considered is defined by an energy
function $V\left(\bf x\right):\mathbb{R}^{3N}\rightarrow\mathbb{R}$,
where $N$ is the number of atoms in the system with a total of $3N$
degrees of freedom (DOF). The potential function is assumed to be
smooth. The force at a point ${\bf
  x}\in\mathbb{R}^{3N}$ is ${\bf F}\left(\bf x\right) = -\nabla
V\left(\bf x\right)$ and the Hessian is given by ${\bf H}\left(\bf
  x\right) = \nabla^{2} V\left(\bf x\right)$. 
The equations that govern the gentlest ascent dynamics are as follows:
\begin{equation}
  \begin{split}
    \bf \dot{x} &= \bf F\left(\bf x\right) - 2\left({\bf F}\left(\bf
        x\right),\bf n\right)n \\ 
    \bf \gamma\dot{n} &= \bf -Hn + \left(\bf n, Hn\right)n
    \label{EqMotion1}
  \end{split}
\end{equation}
The first equation means that we reverse  the force in the direction
of $\bf n$:
$\tilde{\bf F} = \bf F_{\perp} -
F_{\parallel}$, where $\bf F_{\perp}$ and $F_{\parallel}$,
respectively, are the
components of the local force perpendicular to $\bf n$ and parallel to
$\bf n$ and are given by $\bf
F_{\parallel} = \left(\bf F, n\right)n$,  $\bf F_{\perp} = F - F_{\parallel}$. 
The second equation defines the dynamics of the ascent direction $\bf n$.
The first term on the right hand side ensures that $\bf n$ converges
to an eigenvector associated with the smallest eigenvalue of $\bf H$.
The second term ensures that the length of $\bf n$ is fixed at $1$.
Note that ${\bf n}\left(t = 0\right)$ should be a unit vector. If $\bf
H$ is fixed, then the second equation can be regarded as a continuous
version of the power method for finding the eigenvector corresponding
to the smallest eigenvalue. $\gamma$ is a parameter that determines
the rate of convergence of the direction vector to the lowest
eigen-mode of the Hessian. In the limit $\gamma\rightarrow 0$, the
system evolves by following the minimum eigenmode of the local Hessian. 

It was shown by E et al that the only stable fixed points of GAD are
the index-1 saddle points.$\cite{EX11}$ In many ways,  GAD is the
analog of the steepest decent dynamics: 
\begin{equation}
    \bf \dot{x} = \bf F\left(\bf x\right)
    \label{EqMotion2}
\end{equation}

While the steepest decent dynamics are stuck at the local minima, GAD
gets trapped at the index-1 saddle points. 
For this reason, it is of interest to have molecular dynamics or Langevin
dynamics versions of GAD, which will allow us sample to saddle point regions.
This is the one of main purposes of the current paper.

The equations of motion for the stochastic version of GAD are simply:
\begin{equation}
  \begin{split}
    \bf \dot{x} &= \bf F\left(\bf x\right) - 2\left({\bf
        F}\left(\bf x\right),\bf n\right)n + \epsilon\;\eta\left(t\right)\\
    \bf \gamma\dot{n} &= \bf -Hn + \left(\bf n, Hn\right)n
  \label{EqMotion3}
  \end{split}
\end{equation}
where, $\eta\left(t\right)$ is the noise term with variance given by
$\langle\eta\left(t\right)\eta\left(t'\right)\rangle =
\delta\left(t-t'\right)$ and mean value
$\langle\eta\left(t\right)\rangle = 0$. Since a system following
($\ref{EqMotion1}$) gets trapped at an index-1 saddle points, presence of a
noise term in ($\ref{EqMotion3}$) helps the system to hop between
different saddle points.

The dynamical system in ($\ref{EqMotion1}$) can be extended to
molecular dynamics at finite temperature conditions as:
\begin{equation}
  \begin{split}
    \bf \dot{x} &= \bf v\\
    \bf \dot{v} &= m^{-1}\left[\bf F\left(\bf x\right) - 2\left({\bf
        F}\left(\bf x\right),\bf n\right)n\right]\\
    \bf \gamma\dot{n} &= \bf -Hn + \left(\bf n, Hn\right)n
  \end{split}
  \label{EqMotion4}
%  \label{FiniteTempGAD1}
\end{equation}
Similar to GAD, for the set of dynamical equations
($\ref{EqMotion4}$), index-1 saddle points of $V\left(\bf x\right)$
become linearly stable and locally stable fixed points of $V\left(\bf
  x\right)$ become linearly unstable fixed points of
($\ref{EqMotion4}$). A detailed mathematical analysis of the local
convergence is presented in Appendix I. The dynamical system defined by
($\ref{EqMotion4}$) is similar to molecular dynamics, where the system
hops between different basins of attractions around metastable fixed
points of $V\left(\bf x\right)$, except that in ($\ref{EqMotion4}$)
the basins of attractions are index-1 saddle points of $V\left(\bf
  x\right)$. The molecular dynamics version of GAD an also be extended
to perform constant temperature dynamics: 
\begin{equation}
  \begin{split}
    \bf \dot{x} &= \bf v\\
    \bf \dot{v} &= m^{-1}\left[\bf F\left(\bf x\right) - 2\left({\bf
        F}\left(\bf x\right),\bf n\right)n\right] - \beta\bf v\\
    \bf \gamma\dot{n} &= \bf -Hn + \left(\bf n, Hn\right)n
  \end{split}
  \label{EqMotion5}
\end{equation}
where, $\beta$ is a temperature controller similar to Nos$\rm\acute{e}$-Hover
thermostat used in molecular dynamics.$\cite{Nose84, Hoover85}$

It is often desirable to perform simulations under constant external
stress conditions and obtain the ensuing free energy barrier
corresponding to a rare event. To this end, we present a variation of 
MD-GAD based on Parrinello-Rahman extension of molecular dynamics to
simulate a system under constant external stress $\bf S$ and external
pressure $p$.$\cite{ParrinelloR80, ParrinelloR81}$ Let ${\bf h}_{0}$
be the matrix formed by the three initial lattice vectors $\bf a$,
$\bf b$ and $\bf c$, i.e. ${\bf h}_{0} = \{\bf a, b, c\}$ and
$\Omega_{0} = \det\left({\bf h}_{0}\right)$. Correspondingly, the
position vectors $\bf x$ can be represented in terms of reduced coordinates
${\bf x}_{r}$ by the transformation ${\bf x}_{r} = {\bf h}^{-1}{\bf
  x}$, where $\bf h$ is the time dependent metric tensor. The
equations of motion are
\begin{equation}
  \begin{split}
    {\bf \dot{x}}_{r} &= {\bf v}_{r}\\
    {\bf \dot{v}}_{r} &= {\bf h}^{-1}\left[\bf F - 2\left({\bf F},\bf
        n\right)n\right] - {\bf G}^{-1}{\dot{\bf G}{\bf v}_{r}}\\
    W\ddot{\bf h} &= \left({\bf\Pi}-p\right)\sigma - \bf h\Sigma\\
    \bf \gamma\dot{n} &= \bf -Hn + \left(\bf n, Hn\right)n
  \end{split}
  \label{EqMotion6}
\end{equation}
In the above, $\sigma=\partial\Omega/\partial{\bf h}$, $\bf G = h^{T}h$ and $W$
determines the rate at which the system approaches equilibrium between
the external and internal stress.$\cite{ParrinelloR81}$ Descriptions
for $\bf\Sigma$ and $\bf\Pi$ can be found in Ref[19].

\subsection{Illustrative 2-dimensional example}
\begin{figure}[thbp]
   \centering
   \includegraphics[width=0.32\textheight]{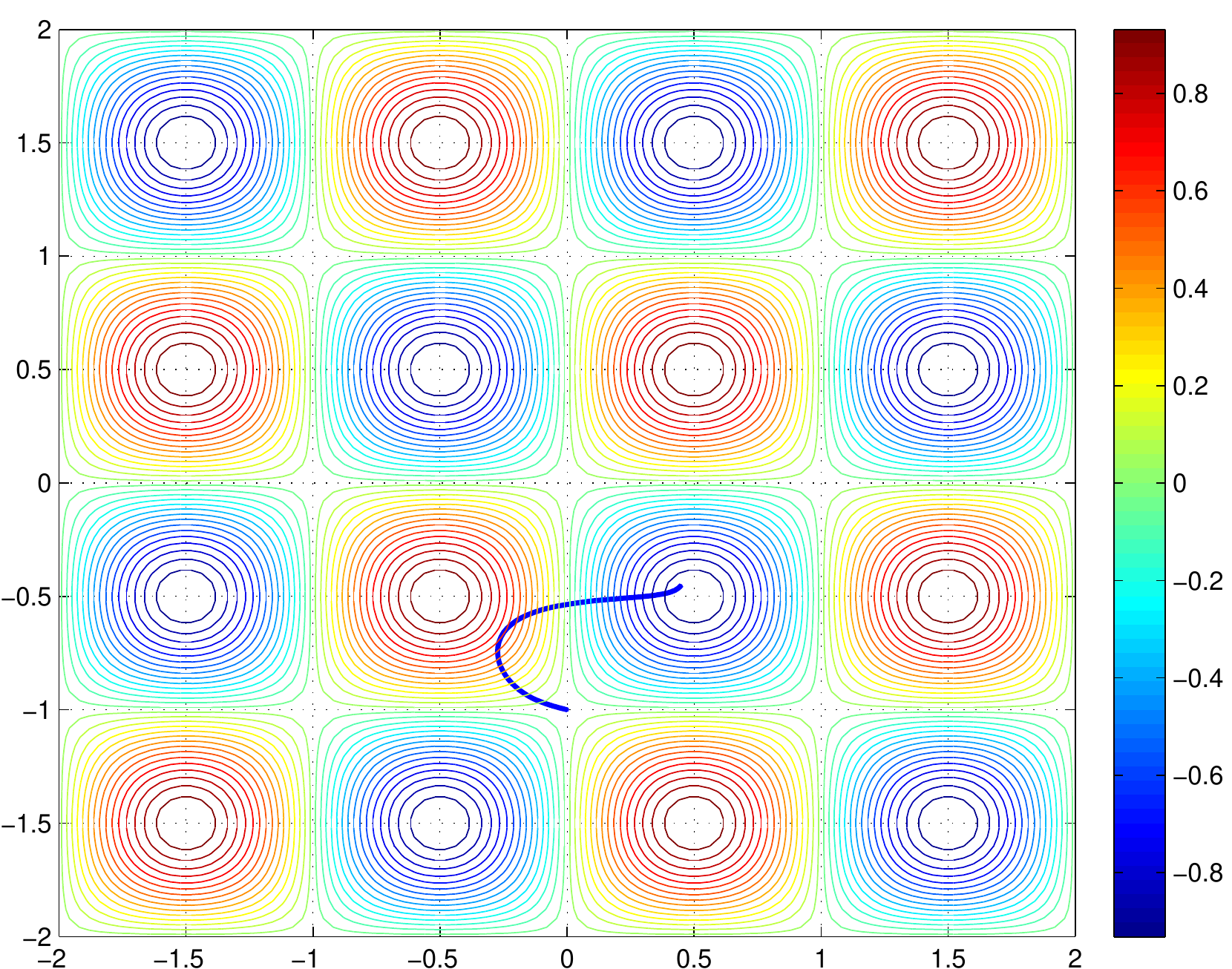}
   \caption{Potential energy surface for the
     2-dimensional toy problem showing the evolution (solid blue line)
     of the system from a metastable state to an index-1 saddle
     point by following ($\ref{EqMotion1}$).}
   \label{PES2dGAD1}
 \end{figure}

We start with a 2-dimensional toy example with the energy function :
\begin{equation}
V\left(x,y\right) = \sin\left(\pi x\right)\sin\left(\pi
  y\right),\qquad x,y\in [-1, 1].
\label{Pote2D}
\end{equation}
$V\left(x,y\right)$ has many locally stable points, such as ($-\frac{1}{2},
\frac{1}{2}$), ($\frac{1}{2},-\frac{1}{2}$), etc. and saddle points
($0,0$), ($\pm 1, 0$), etc. inside the domain of interest. First set
of simulations are performed using the deterministic version of GAD
prescribed in ($\ref{EqMotion1}$). The simulation is started at a
point close to the metastable fixed point ($\frac{1}{2},
-\frac{1}{2}$) with a randomly initialized direction
vector. Fig. $\ref{PES2dGAD1}$ shows the evolution of the system in 
the two dimensional configuration space. As the system moves out of
the energy well, the direction vector 
slowly relaxes to the smallest eigenvalue of the Hessian and then
converges to the saddle point at (-1, 0). An important conclusion that can be drawn
from these results is that the path can deviate significantly from the
minimum energy path and the choice of the initial direction vector
determines the location of the converged saddle point. Consequently
different realizations starting from the same initial point can
converge to different saddle points.

\begin{figure}[thbp]
   \centering
   \subfigure[]  {
     \includegraphics[width=0.32\textheight]{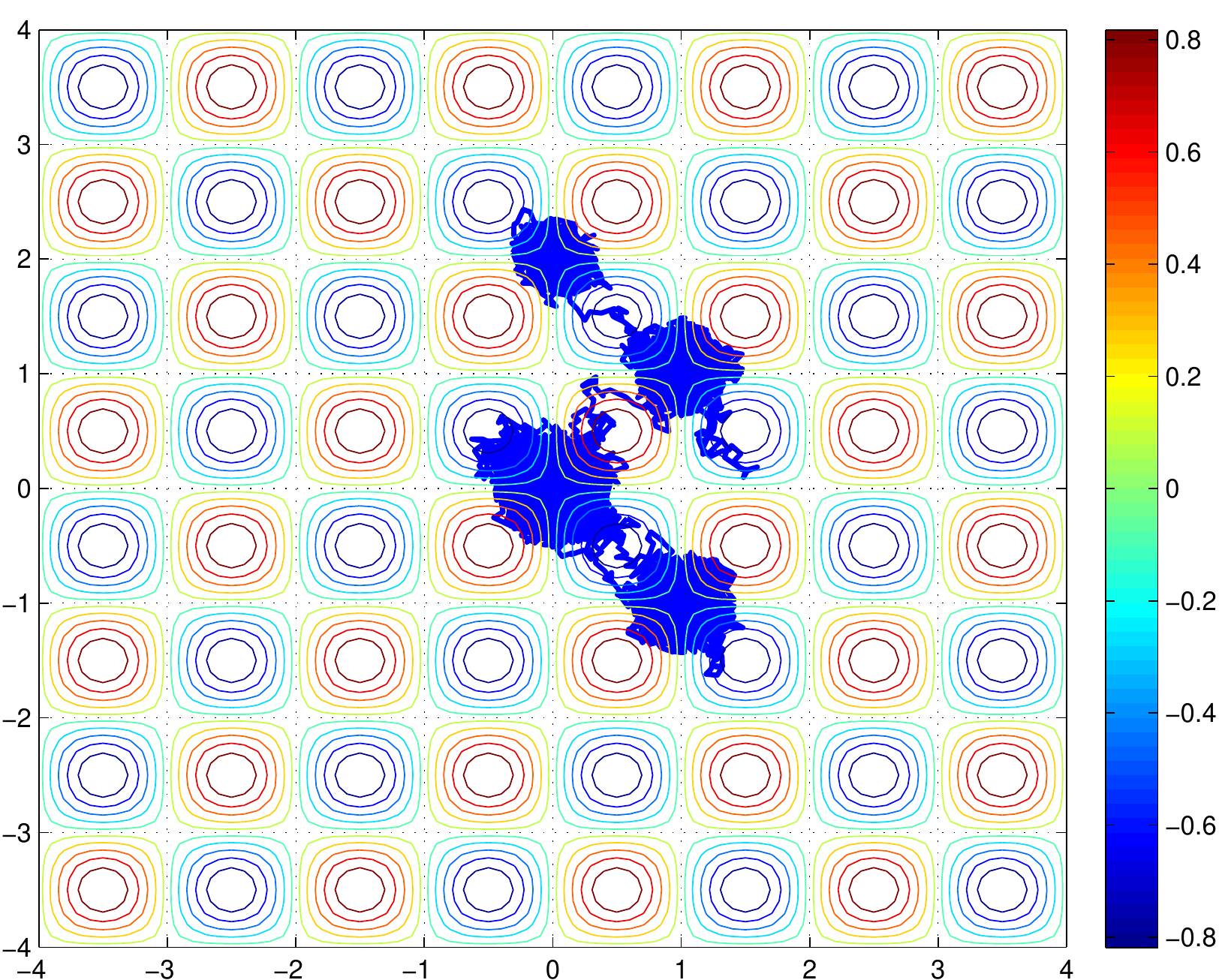}
     \label{PES2dGADnoise1}
   }
   \subfigure[] {
     \includegraphics[width=0.335\textheight]{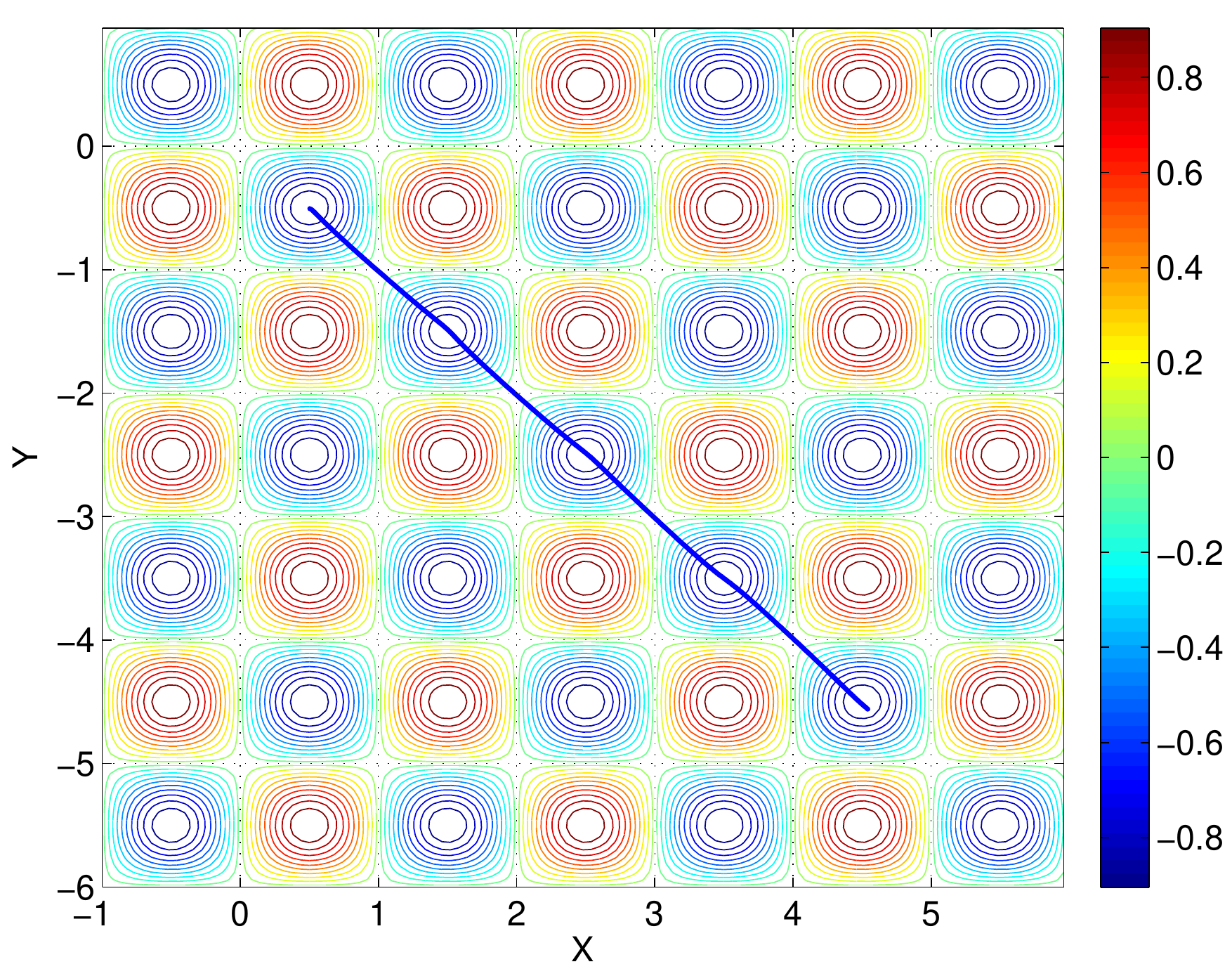}
     \label{Energy2dGADnoise2}
     }
   \caption{$\ref{PES2dGADnoise1}$ Solutions of ($\ref{EqMotion3}$)
     showing the trajectory followed by a system on the 2-dimensional
     PES. The system spends majority of its time near saddle
     points. $\ref{Energy2dGADnoise2}$ Shows the evolution of the
     system using MD-GAD prescribed in ($\ref{EqMotion4}$). In this
     case the system can easily traverse through many locally stable
     and index-1 saddle points of $V\left(x,y\right)$.}
 \end{figure}

The stochastic version of GAD can be used to navigate the PES to sample
different saddle points. Fig. $\ref{PES2dGADnoise1}$ shows the results
for stochastic dynamics performed by the system on the PES. The system
spends majority of its time around saddle points in contrast to
general overdamped Langevin dynamics in which the system samples
stable fixed points of the PES. 
 
Next, we turn to MD-GAD prescribed in ($\ref{EqMotion4}$). In
Fig. $\ref{Energy2dGADnoise2}$ the system travels from the initial
locally stable fixed point at ($\frac{1}{2}, -\frac{1}{2}$) to the
final metastable state at ($\frac{9}{2}, -\frac{9}{2}$) by passing
through many saddle points and locally stable fixed points. The
simulation is started with zero kinetic energy. As the system walks
out of the initial energy well and moves towards the saddle point, the
KE of the system increases and attains a maximum value. This pushes
the system towards the next metastable point where the KE reduces to
zero. Subsequently, the system again gains kinetic energy as it moves
close to the next saddle point. As this process is repeated many fixed
points of the PES are sampled. 

\section{Applications}
To further illustrate our approach, we simulate an ensemble of saddle
points for two systems: (a) a heptamer island on $(111)$ surface of copper
and (b) a surface vacancy and ad-atom pair on the $(111)$ surface of copper. The
heptamer island system has been widely studied by Henkelman et
al. using the dimer method and most recently by J$\rm\ddot{o}$nsson et al. to
benchmark the convergence of different saddle point finding
strategies.$\cite{HenkelmanJ99, MironF01, PedersenHJ11}$ The second
example has clear time-scale separation between different rare events,
namely, ad-atom diffusion and vacancy diffusion and hence presents
excellent opportunity to test the scope of saddle point finding
algorithms. While diffusion of an ad-atom on a metal surface has been
studied before the presence of a vacancy adds significant complexity
to the problem.$\cite{MironF01, PedersenHJ11}$ Our intention is to
explore the high dimensional PES close to these locally stable
configurations to probe and catalogue some of the probable surface
activity that can play an important role in other scenarios where
quantitative analysis of different rare events may be difficult. 

\subsection{Selecting an ``active region'' of the sample}
The problem of determining saddle points close to a local minimum
entails moving out of the potential energy well along
preferential directions in the multidimensional configuration
space. The challenge in sampling these preferential directions comes
from the fact that they represent only a miniscule
fraction of all the possible directions emanating from the initial
locally stable fixed point on the high dimensional PES. 

Our aim is to significantly reduce the size of the configuration
space by selecting a small, yet important, region of the sample. In
materials physics, many deformation processes can be easily traced by
analyzing the softest eigenmodes of the Hessian.
This helps in determining a vulnerable region of the system. In
systems where one has an intuitive understanding of the vulnerable
region ($\Omega_{v}$) one can select a set of atoms and include their
first nearest neighbors to form a set of ``active region'' $\Omega$,
such that $\Omega_{v}\subset\Omega$. We
show that a more prudent way to sample $\Omega_{v}$, which is prone to
deformation, is by analyzing the unit force vector close to the initial 
locally stable configuration. Once the vulnerable DOF are identified,
the corresponding atoms and their first nearest neighbors are selected
to form the ``active region'' of the sample. Our assumption is that
this ``active region'' corresponds to a significantly reduced
configuration space where probable low barrier transition events can
take place. 

\begin{figure}[thbp]
   \centering
   \subfigure[] {
     \includegraphics[width=0.45\textheight]{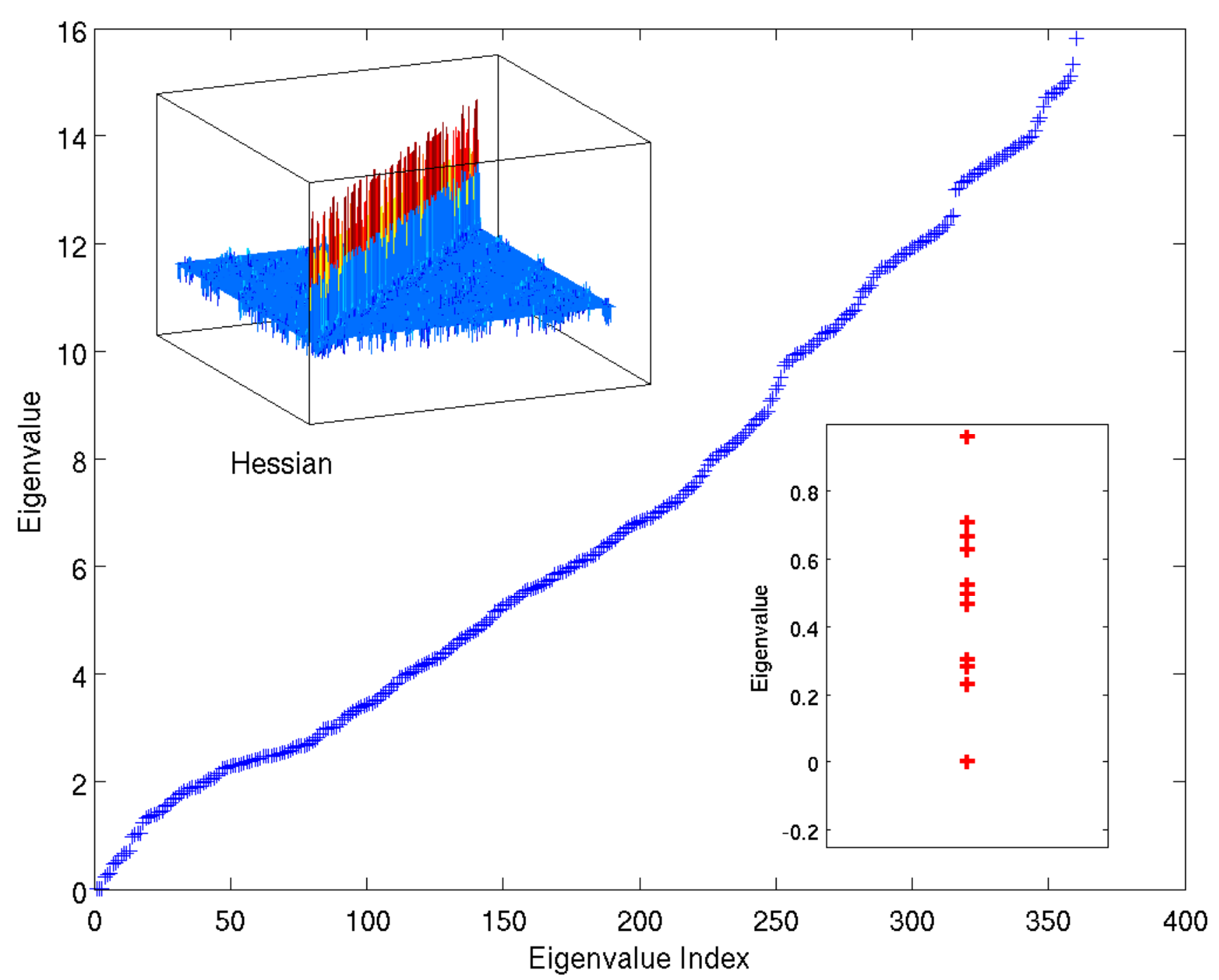}
     \label{EigValHessianVacAdatom}
     }
   \subfigure[]  {
     \includegraphics[width=0.45\textheight]{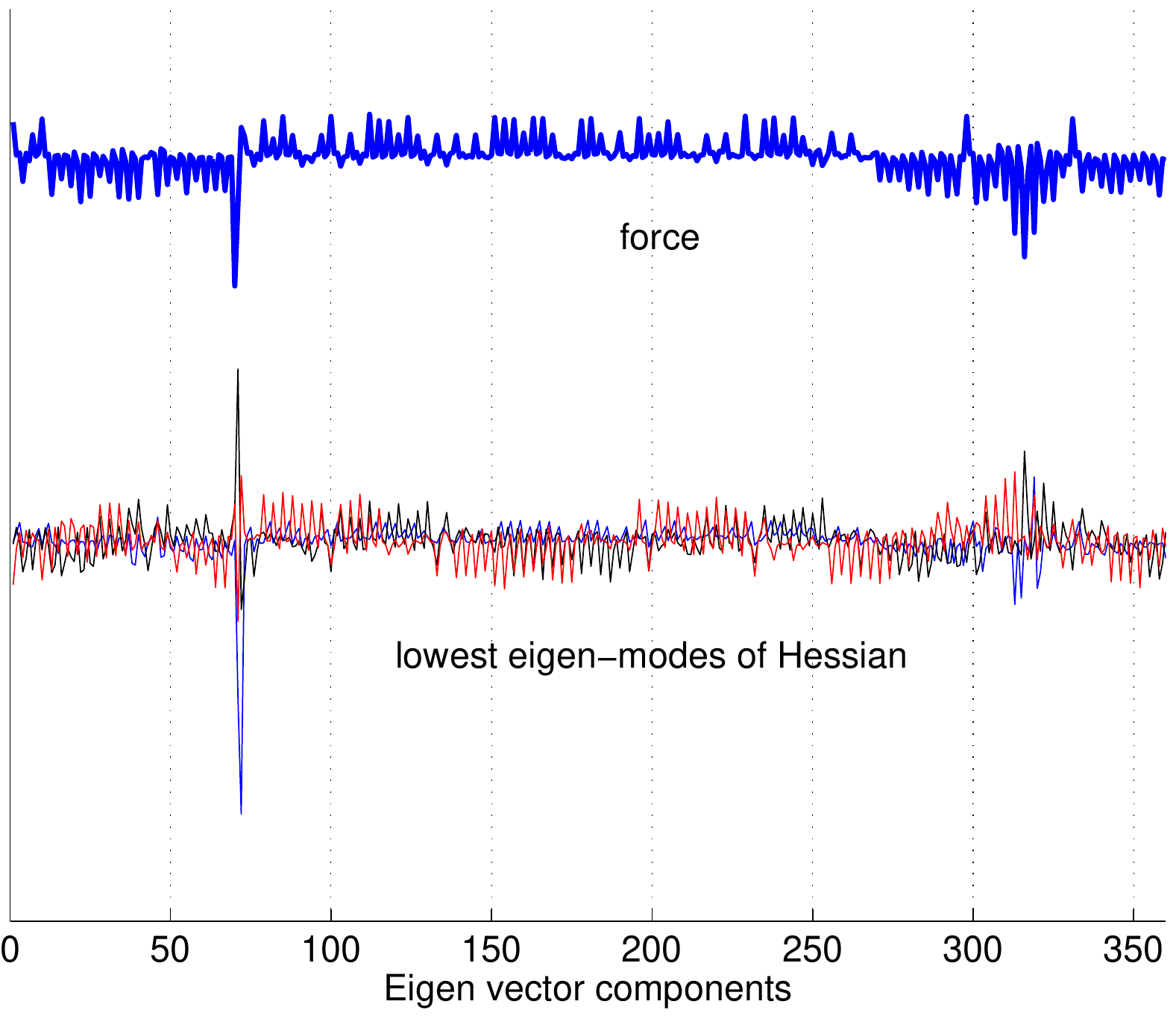}
     \label{EigVecHessianVacAdatom}
   }
   \caption{$\ref{EigValHessianVacAdatom}$ Shows the eigenvalues of the
     Hessian, arranged in ascending order, at the local minimum
     configuration for the vacancy ad-atom system. The upper inset
     shows that the Hessian is diagonal dominant and the lower inset
     shows the clustering of low lying eigenvalues of the
     Hessian. $\ref{EigVecHessianVacAdatom}$ Shows the different DOF
     of the unit force vector (blue) and the superposition of 6 lowest
     eigenvectors of the Hessian at the local minimum
     configuration. The vulnerable DOF can be easily isolated.} 
 \end{figure}

Let us consider the system containing a surface vacancy and an ad-atom
on the $(111)$ surface of a copper thin film
(Fig. $\ref{AdatomLocMin}$). For this system, we have explicitly
evaluated Hessian for the relaxed initial configuration (force norm $<
10^{-4}$ eV/$\rm\AA$). The elements in the $i$-th row of the Hessian
matrix are determined by displacing the $i$-th DOF by a small
displacement $\delta x_{i}$ from the equilibrium structure and
evaluating the change in the force on all atoms. The Hessian
(characterized by the upper inset in
Fig. $\ref{EigValHessianVacAdatom}$) is then diagonalized and its
eigenvalues calculated. Fig. $\ref{EigValHessianVacAdatom}$ shows the
spectrum of the Hessian of the system. The lower inset in the figure
shows the clustering of low-lying eigenvalues which may be difficult
to resolve by using power method based tools. A superposition of the
low-lying eigenvectors (first 6 lowest eigenvectors, neglecting the
global rotation and global translation components) are plotted in
Fig. $\ref{EigVecHessianVacAdatom}$.  

Fig. $\ref{EigVecHessianVacAdatom}$ also shows the contributions of
the different DOF to the unit force vector (shown in blue) at the initial locally
stable fixed point of the energy landscape. Upon comparison with the
superposition of the six low-lying eigenvectors, we find the the
DOFs that have significantly higher weights in the unit force vector
also correspond to those in the low lying eigenvectors of the Hessian. This suggest
that close to a locally stable fixed point the force can provide
information about the vulnerable DOF of the system and can be used to
obtain probable initial direction vectors for probing saddle points on
the PES. Following the above argument, using the unit normed force at the
local minimum point, we selected a set of 23 atoms as our ``active
region'' $\Omega$.

\begin{figure}[thbp]
   \centering
   \subfigure[]  {
     \includegraphics[width=0.4\textheight]{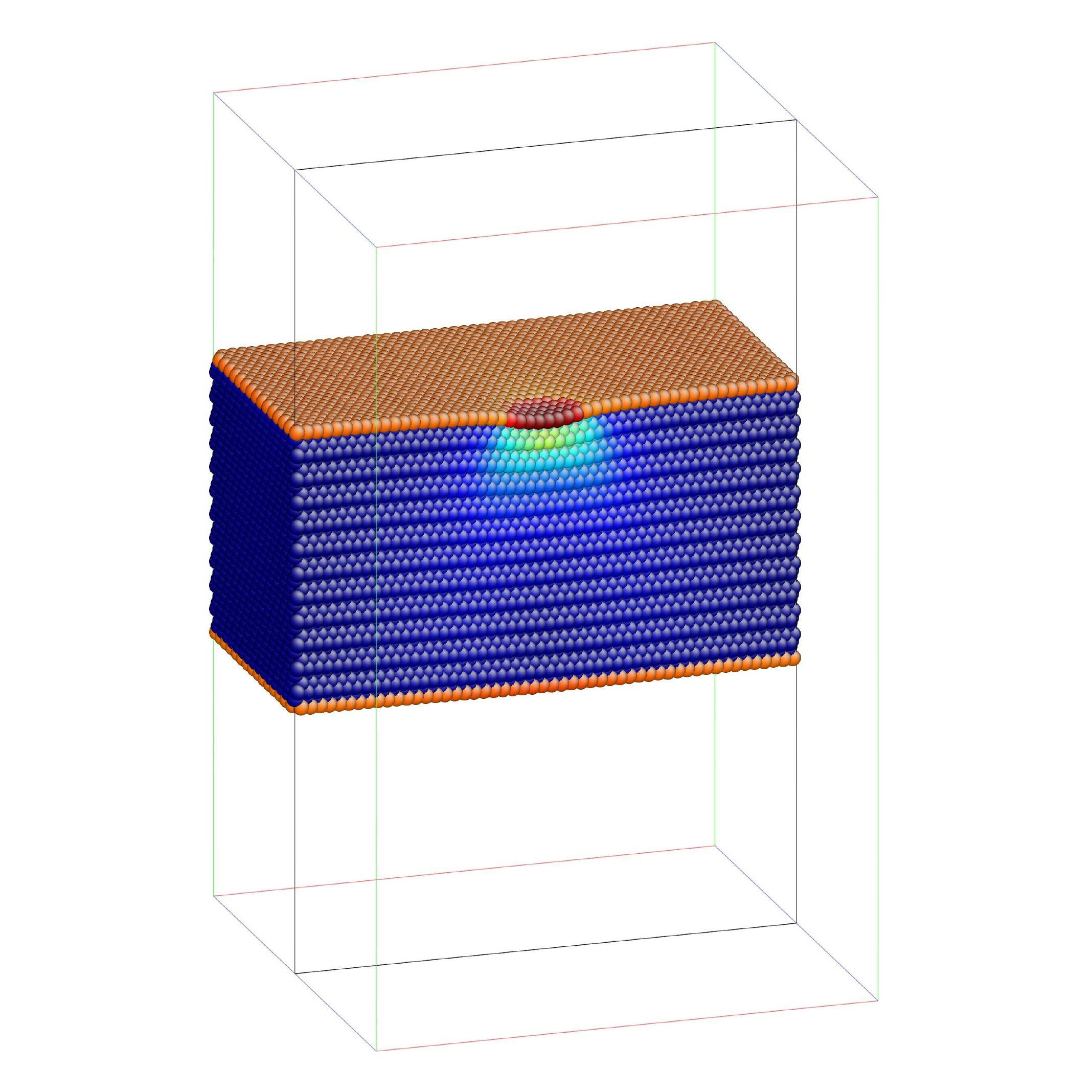}
     \label{NanoIndentation1}
   }
   \subfigure[] {
     \includegraphics[width=0.5\textheight]{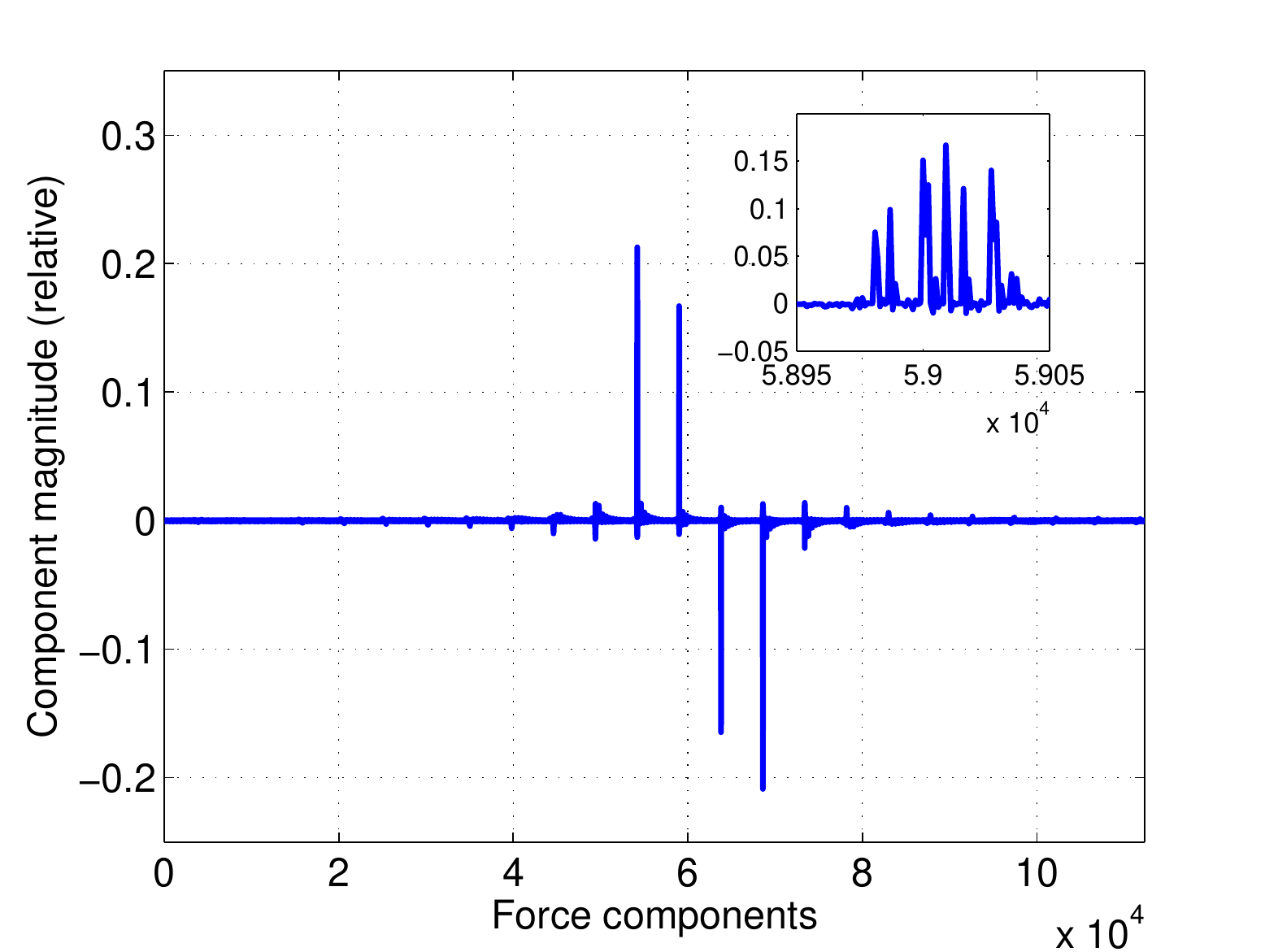}
     \label{NanoIndenationForce1}
   }
   \caption{Simulation cell set-up for nanoindentation on (111)
     surface of an initially defect-free Cu thin-film containing
     112,320 atoms. $\ref{NanoIndentation1}$ shows a cross section of the
     system after indentation performed using a spherical smooth indenter of
     radius 25 $\rm\AA$ on the (111) free surface. The
     indentation direction is perpendicular to the (111) surface. The atoms have 
     been colored by their shear strain (dark blue corresponds to zero strain and
     red corresponds to maximum shear strain $\sim
     0.20$). $\ref{NanoIndenationForce1}$ shows the contributions from
     different DOF to the unit force thus showing the atoms vulnerable
     to deformation. The inset shows a magnified
     portion of the plot. The vulnerable DOF can be easily identified.}
 \end{figure}

We find that even for complicated non-equilibrium processes involving
many potential transition events in presence of external fields (such as
stress, etc.) the unit force on the system can be very
informative. To illustrate this we use the example of
nanoindentation. Fig. $\ref{NanoIndentation1}$ shows a representative
system containing 112,320 copper atoms. The simulation cell has
lattice vectors along $[1\bar{1}0]$, $[111]$ and $[11\bar{2}]$
directions. While free surface conditions are prevalent along $[111]$
direction, periodic boundary conditions are imposed along the
perpendicular directions. The system is elastically indented 
using a smooth spherical indenter (radius 25 $\rm\AA$), imposed by a
force field, to an indentation depth of 4.19 $\rm\AA$. After
indentation, constrained minimization (force norm $< 10^{-2}$
eV/$\rm\AA$) is performed to obtain a locally stable
configuration. Fig. $\ref{NanoIndenationForce1}$ shows the
contributions of the different DOF in the unit force vector from which
we can easily select $\Omega$. Analysis of the probable transition
events taking place during nanoindentation will be presented elsewhere.

\subsection{Initial direction vectors}
Since the overall system size corresponds to a large configuration
space, an activated region of the sample is selected following a
procedure described in the previous section. From the set of atoms
within $\Omega$, we obtain a set of initial direction
vectors by the following procedure : select $m$ atoms and
randomly initializing the 3$m$ direction components while keeping the
remaining components of the direction vector equal to zero. The set of $m$ atoms
is selected by picking an atom $i$ and its ($m-1$)
nearest neighbors from $\Omega$. If the atom $i$ has $n$
($<m-1$) nearest neighbors inside $\Omega$ then the direction has only ($3n+3$) randomly
initialized components and the remaining DOF are set to zero. For our
study of rare events in the representative systems described earlier
we use $m$ = 1, 2, 4, 12, though one can resort to a different set of
values. 

Our choice of selectively activating only a few components of the
direction vector stems from the fact that even though physical
processes, such as different deformation mechanisms, taking place in
most materials have activation volumes in the range of
0.1$b^{3}$-1000$b^{3}$, $b$ is the Burgers vector of the system, the
size of the nucleus in the initial stages can be quite 
small. A smaller activation volume corresponds to a localized process
involving only few atoms such as
interstitial diffusion, vacancy diffusion etc., while a larger
activation volume corresponds to a process which is more delocalized,
such as, forest hardening, Orowan loopin, etc.$\cite{DaoLADM07}$ In each of these
processes, since only a few degrees of freedom of the direction vector
are explored, ideally one needs humongous number of randomly
initialized vectors to sample majority of the most probable rare events. Our
proposed selection procedure overcomes this sampling problem.

\subsection{Convergence criteria}
Once a set of the desired direction vectors is obtained, one can move
out of basins of attraction by moving along these directions. GAD
guarantees that the system converges to an index-1 saddle point on the
energy surface. However, the convergence will depend on how fast the
system can relax to the lowest eigenmode in the vicinity of the saddle
point. During this time it is possible that the system might have
traversed through few different fixed points on the PES. In order to
avoid this ``overshooting'' one might resort to a smaller time
step. However, this is not conducive as the majority of initialized
direction vectors have sufficient time to relax to the minimum
eigenmode much before the system reaches the separatrix. 

Hence, a pragmatic approach is needed to determine whether the system
has left the basin of attraction or not. Our approach is motivated by the
existence of an unique equilibrium bond length ($r_{\rm o}$) between nearest
neighbor atoms in a locally stable configuration. A rare event
involves a process of bond-breaking between nearest neighbors (which
corresponds to a bond length, $r$) in the
group of atoms taking part in the process. This criteria can be used
to identify the approach of a system to a saddle point. We use the ratio $q =
\left(r-r_{\rm o}\right)/r_{\rm o}$ for this purpose. To sample saddle
points close to a local minimum, suggested
values of $q_{c}$ lie in the range $\sim 0.30-0.40$.$\cite{MironF03, MironF04}$
Using this criteria has an added flexibility : one can selectively
tune $q_{c}$ to determine a distribution of saddle points within a
given distance from the initial configuration in the hyper-space.

For the sampling saddle points in the representative
examples we use $q_{c}$ = 0.50. The system follows ($\ref{EqMotion1}$)
with $\gamma=1$ for $q\le q_{c}$. To identify any bond breaking event,
$q$ is calculated every 10 steps. Once, $q>q_{c}$ we use $\gamma = 0.25$
and for each evolution step of $\bf x$, the direction vector $\bf n$
is evolved 2 times to converge to the minimum eigenmode. 

\subsection{Surface vacancy ad-atom pair on (111) surface of cooper}

\begin{figure}[thbp]
   \centering
   \includegraphics[width=0.35\textheight]{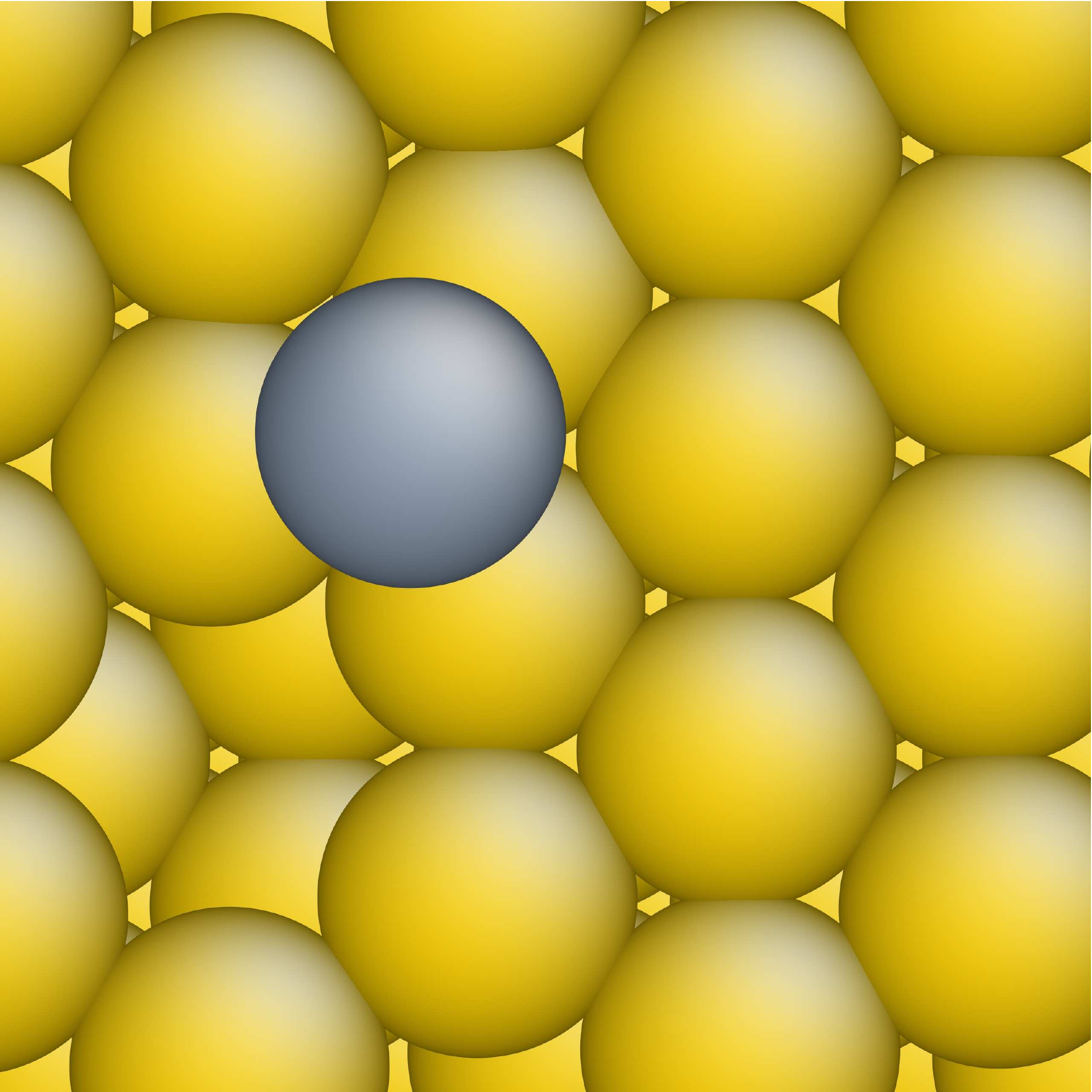}
   \caption{Local minimum configuration for the system containing a
     surface vacancy and an ad-atom (shown in grey) on (111) surface
     of a copper thin film.}
   \label{AdatomLocMin}
 \end{figure}

The simulation setup consists of a, 480 Cu atoms, simulation cell with
axis vectors along $[110]$, $[1\bar{1}1]$ and $[\bar{1}12]$
directions. The set up consists of 6 $(111)$ planes stacked on top of
each other with periodic boundary conditions imposed along $[110]$
and $[\bar{1}12]$ directions while free surface conditions maintained
along $[1\bar{1}1]$ direction. On atom is removed from
the surface to form a surface vacancy and placed on a vacant fcc site
close to the vacancy to form an ad-atom. Fig. $\ref{AdatomLocMin}$
shows the initial configuration obtained by local minimization to a
force of 10$^{-4}$ eV/$\rm\AA$. The ad-atom is shown in
grey. Fig. $\ref{EigVecHessianVacAdatom}$ shows the weights of
different DOF in the unit normed force vector for the initial
configuration. The vulnerable DOF are easily identifiable. These DOF
are used to generate 661 initial direction
vectors. 

\begin{figure}[thbp]
   \centering
   \subfigure[]  {
     \includegraphics[width=0.24\textheight]{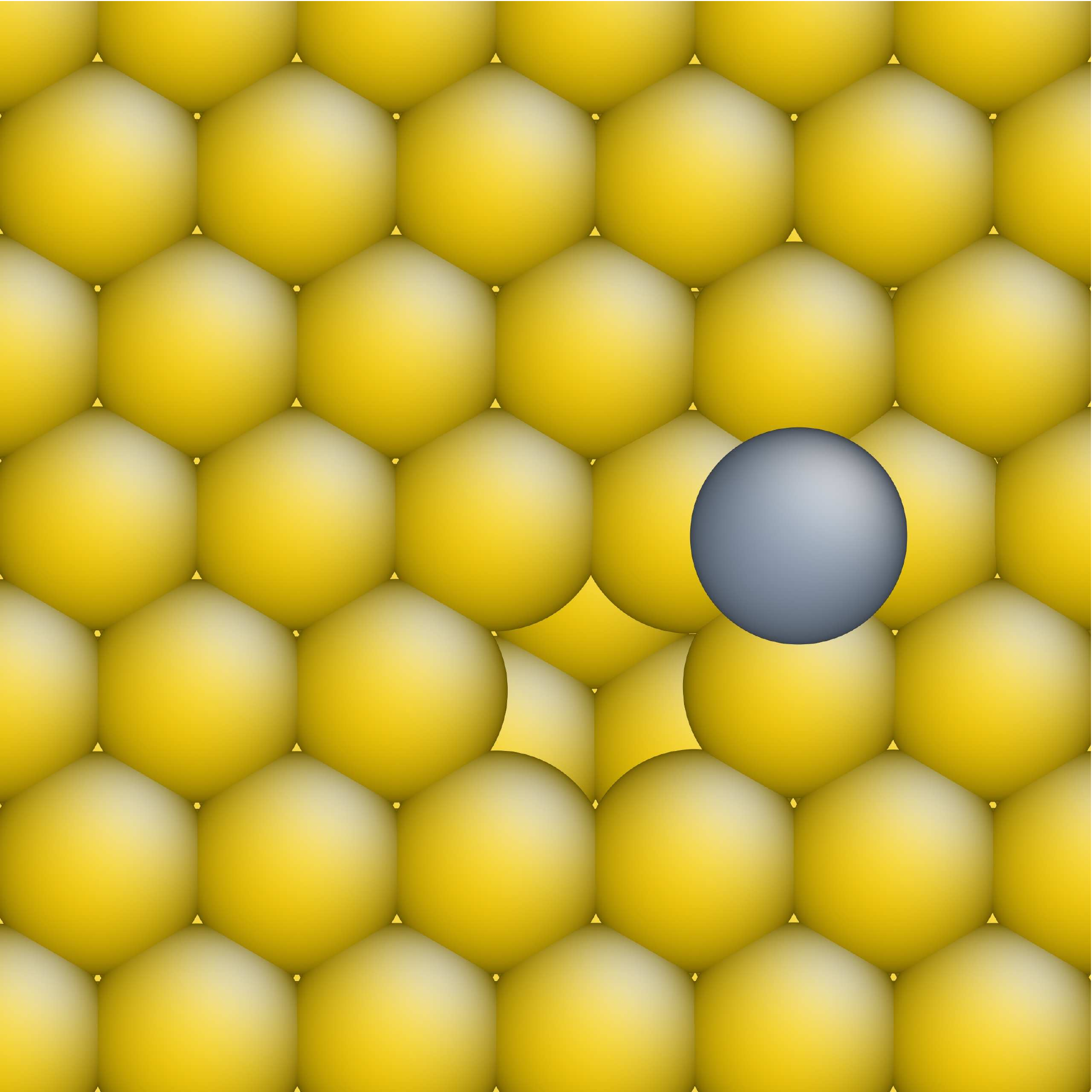}
     \label{VacAdatom144}
   }
   \subfigure[] {
     \includegraphics[width=0.24\textheight]{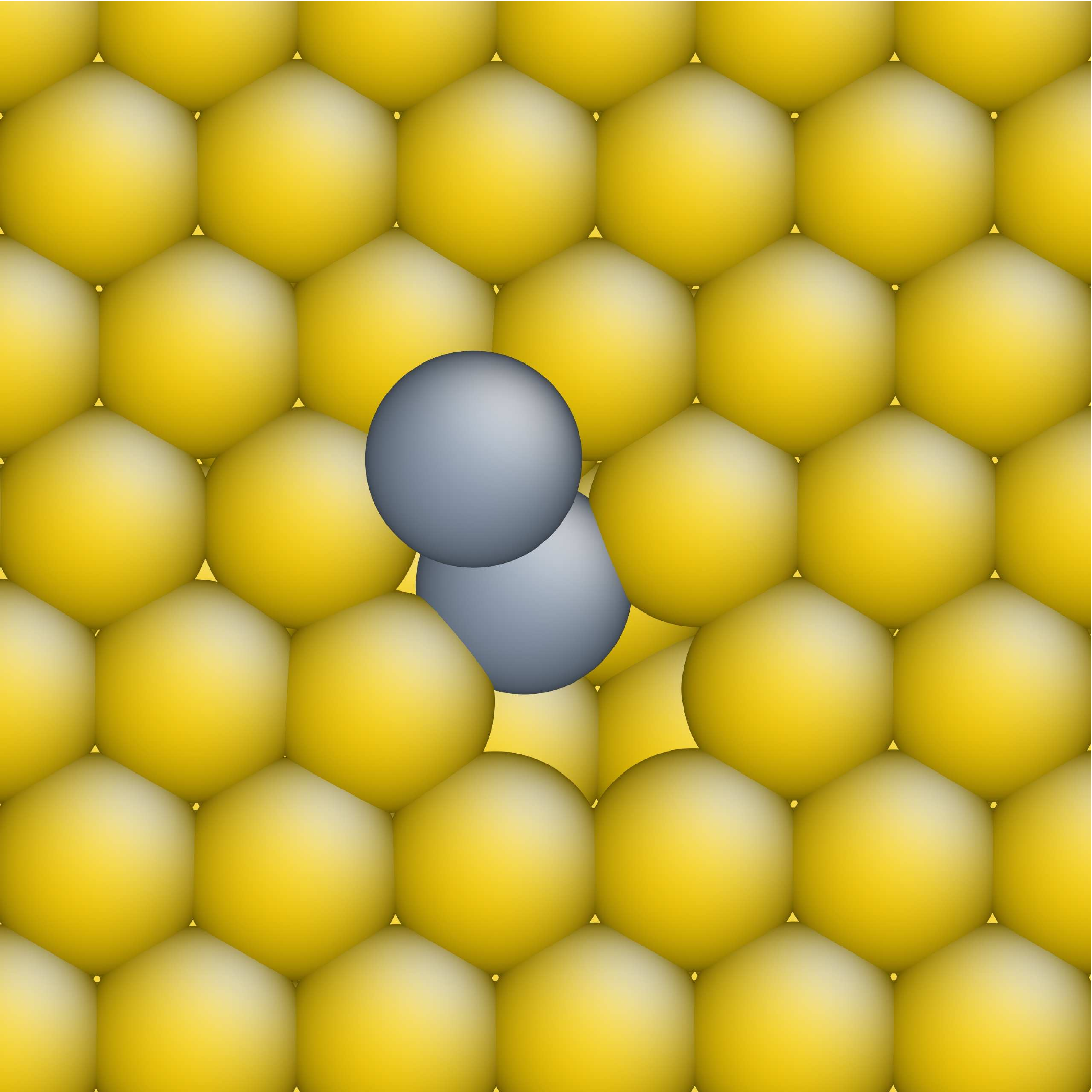}
     \label{VacAdatom128}
   }   
   \subfigure[] {
     \includegraphics[width=0.24\textheight]{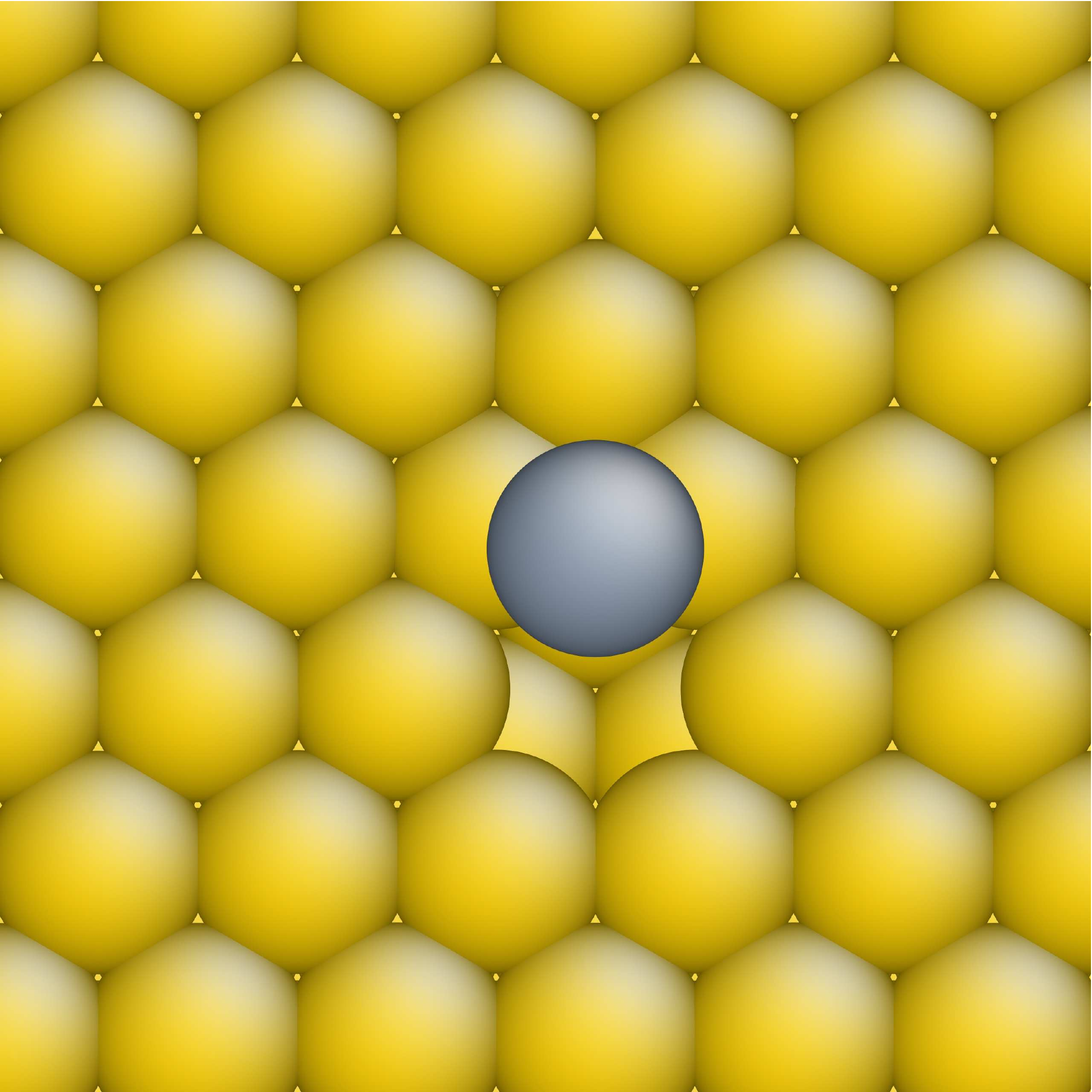}
     \label{VacAdatom233}
   }
   \subfigure[] {
     \includegraphics[width=0.24\textheight]{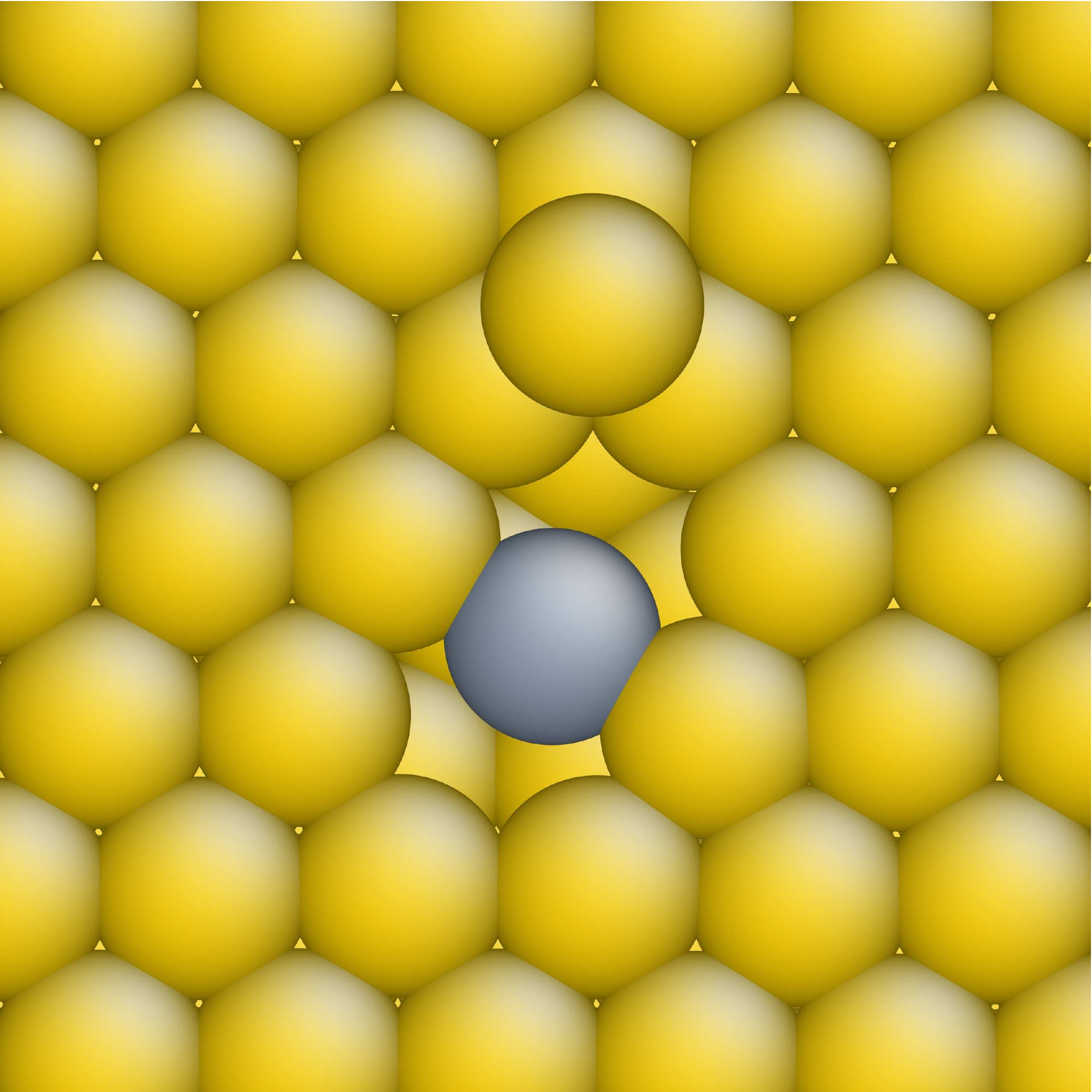}
     \label{VacAdatom11}
   }
   \subfigure[] {
     \includegraphics[width=0.24\textheight]{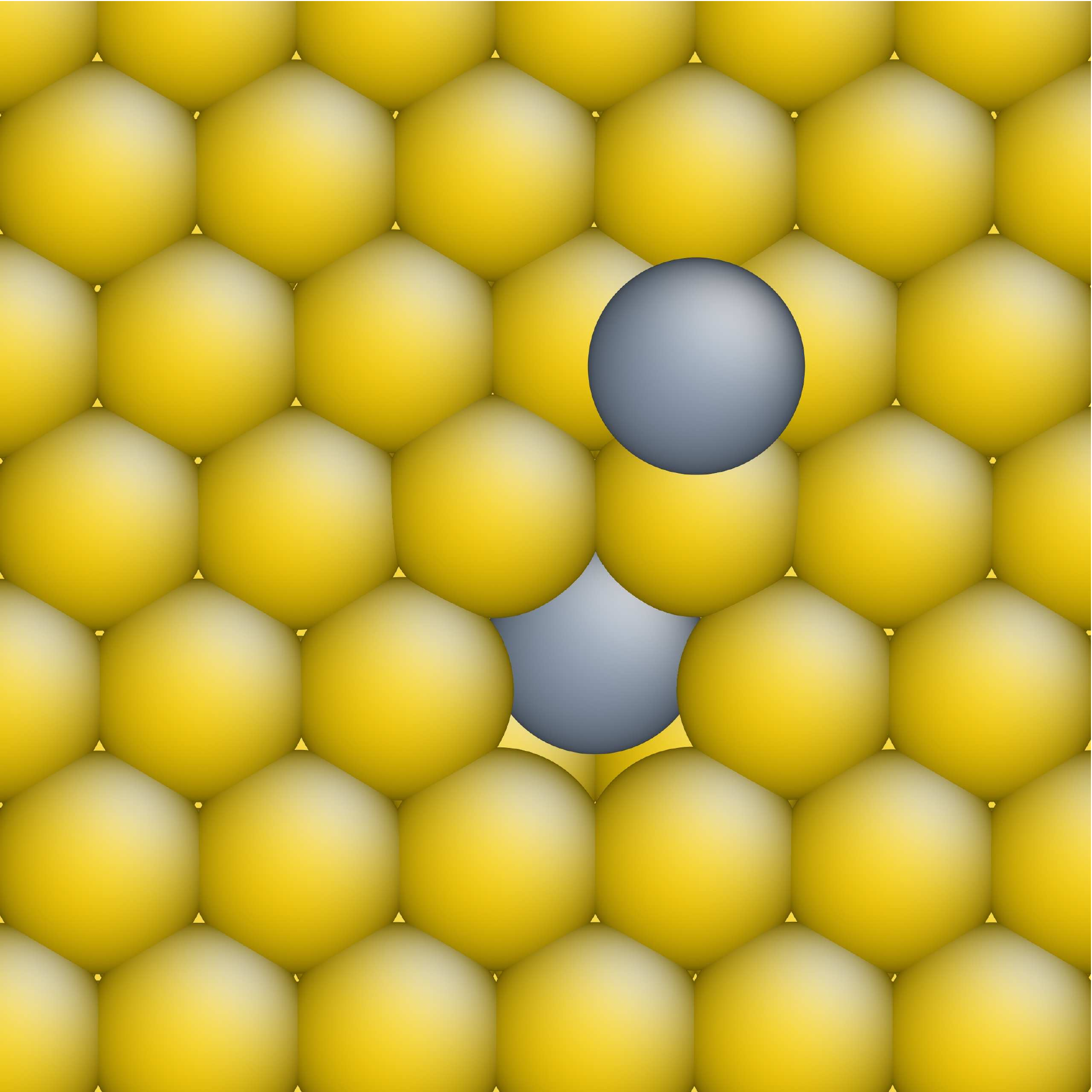}
     \label{VacAdatom52}
   }   
   \subfigure[] {
     \includegraphics[width=0.24\textheight]{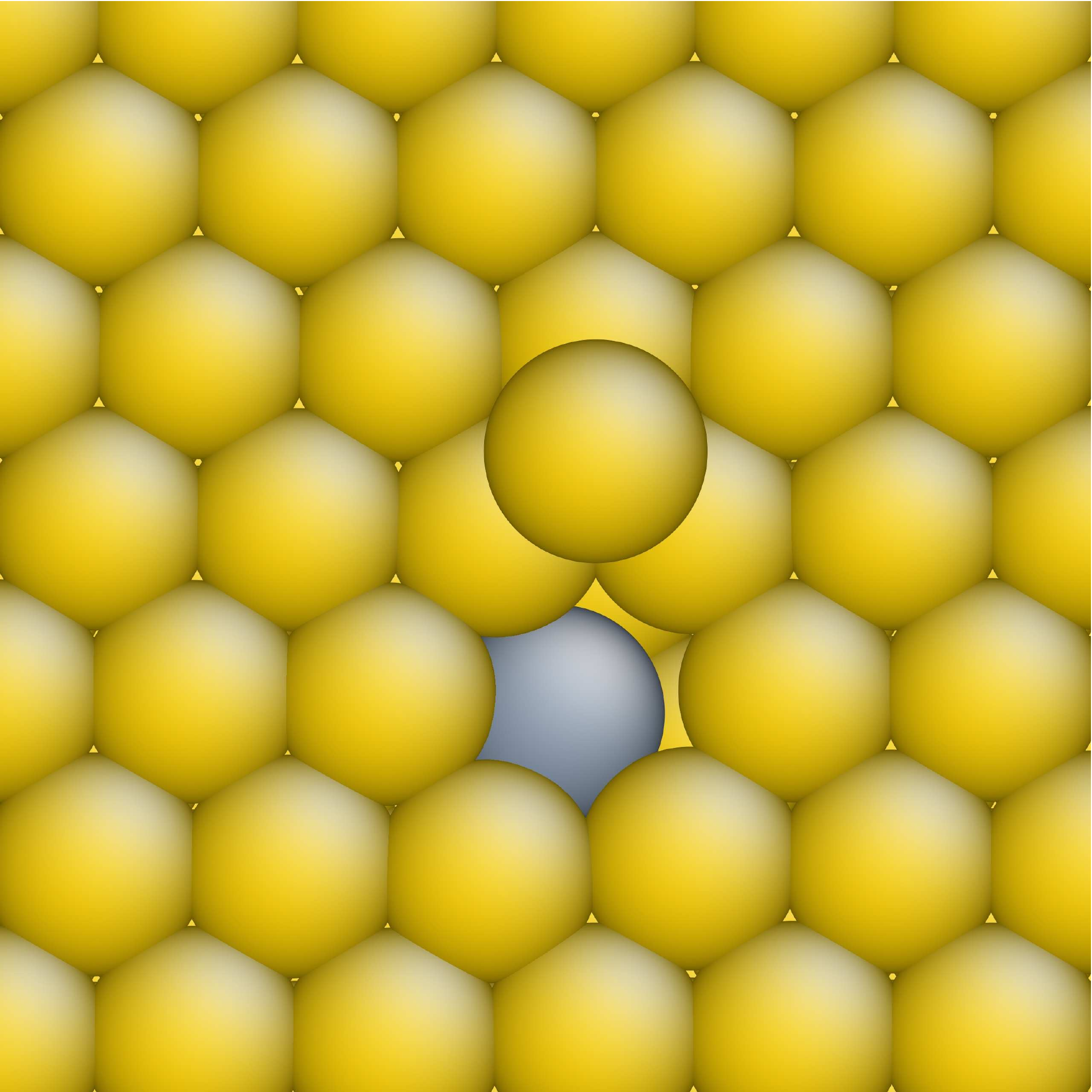}
     \label{VacAdatom13}
   }
   \caption{Some converged saddle configurations for the vacancy ad-atom
     system obtained from GAD simulations. $\ref{VacAdatom144}$
     ad-atom diffusion (barrier 0.01 eV),  $\ref{VacAdatom128}$
     collective process involving ad-atom and vacancy migration
     (barrier 0.15 eV), $\ref{VacAdatom233}$ vacancy ad-atom
     annihilation (barrier 0.30 eV), $\ref{VacAdatom11}$ vacancy
     diffusion (barrier 0.59 eV), $\ref{VacAdatom52}$ sub-surface vacancy
     formation (barrier 0.77 eV), $\ref{VacAdatom13}$ sub-surface vacancy
     formation (barrier 0.82 eV). Some of the structures are shifted
     in plane for proper viewing.} 
 \end{figure}

\begin{figure}[thbp]
   \centering
   \subfigure[] {
     \includegraphics[width=0.24\textheight]{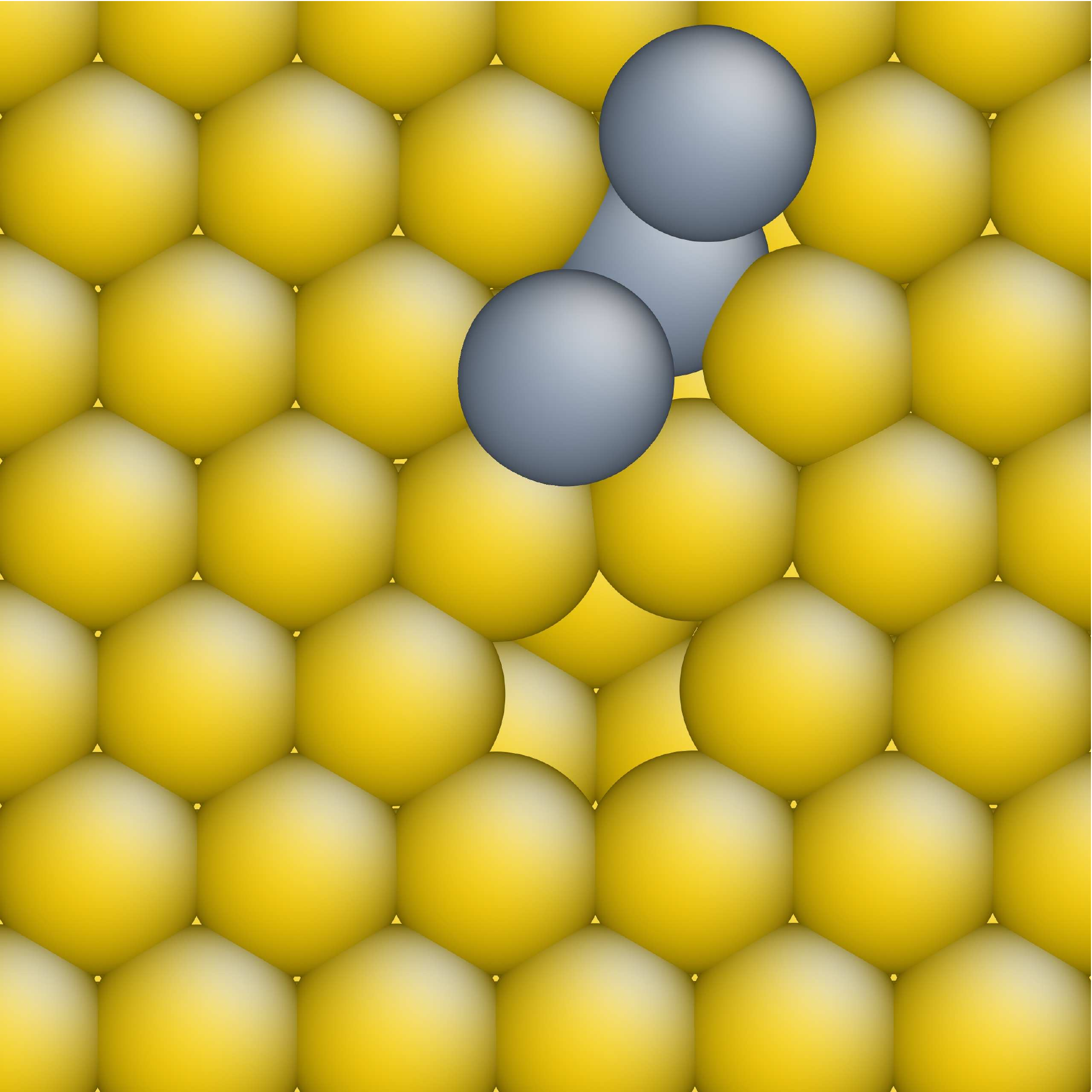}
     \label{VacAdatom345}
   }
   \subfigure[] {
     \includegraphics[width=0.24\textheight]{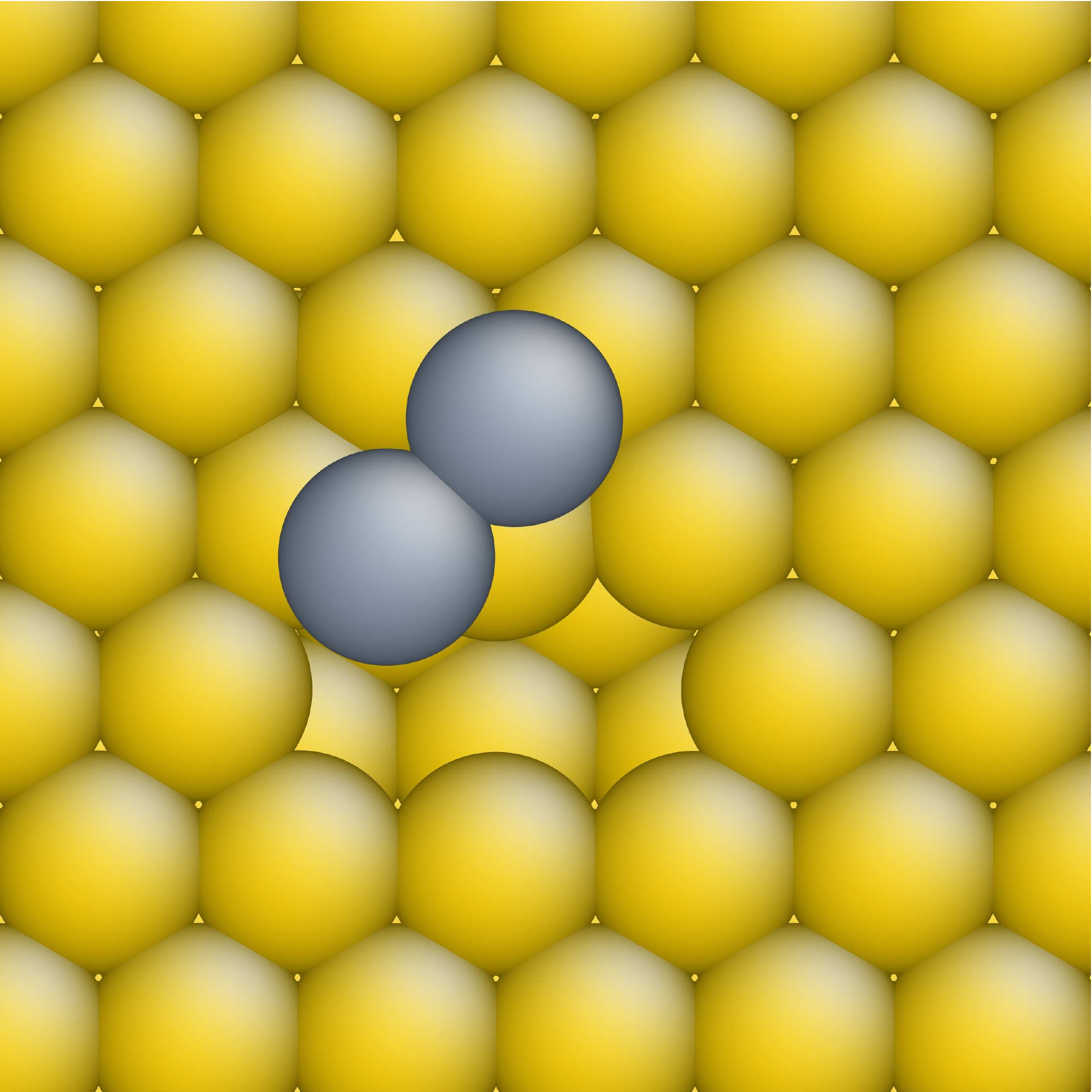}
     \label{VacAdatom157}
   }    
   \subfigure[] {
     \includegraphics[width=0.24\textheight]{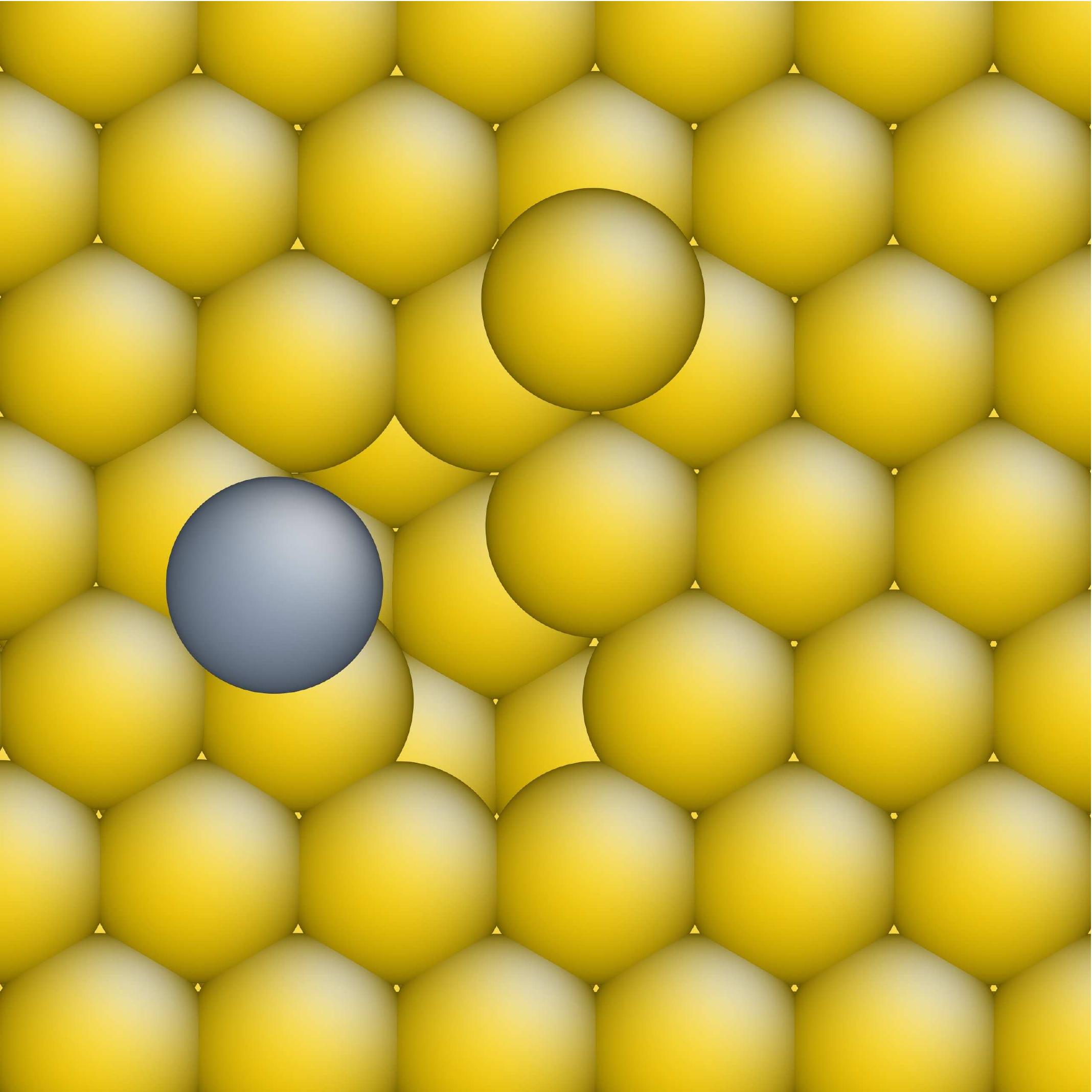}
     \label{VacAdatom601}
   }   
   \subfigure[] {
     \includegraphics[width=0.24\textheight]{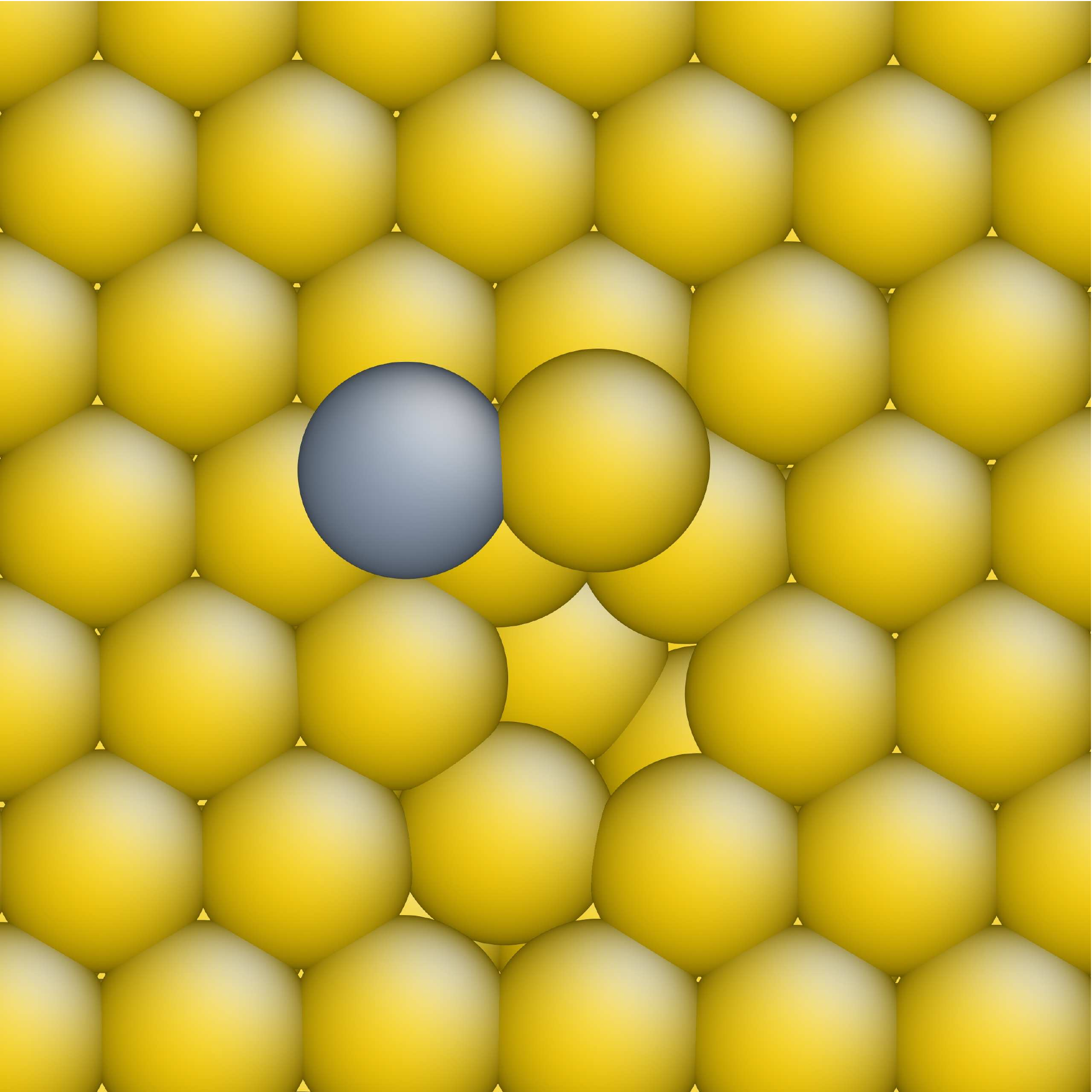}
     \label{VacAdatom100}
   }   
   \caption{Some converged saddle configurations for the vacancy ad-atom
     system obtained from GAD simulations. $\ref{VacAdatom345}$
     ad-atom formation by exchange mechanism (barrier 1.14 eV),
     $\ref{VacAdatom157}$ divacancy and ad-atom formation (barrier
     1.40 eV), $\ref{VacAdatom601}$ divacancy and ad-atom formation (barrier
     1.81 eV), $\ref{VacAdatom100}$ bulk-vacancy and ad-atom formation (barrier
     2.08 eV). }
 \end{figure}

Fig. $\ref{VacAdatom144}$ to Fig. $\ref{VacAdatom100}$ shows the
probable transition events with increasing activation energy
barriers. The smallest barrier (0.01 eV) corresponds to an ad-atom
diffusion. The difference in energy between a fcc and hcp site on a
free surface is much smaller than that in metals like
Pt, resulting in a significantly smaller ad-atom diffusion barrier. We
found the process of diffusion of both ad-atom and vacancy in a
concerted way, as shown in Fig. $\ref{VacAdatom128}$ has a little
higher barrier of 0.15 eV. The other significant events are ad-atom diffusion
to annihilate the ad-atom and surface vacancy pair (Fig. $\ref{VacAdatom233}$),
vacancy diffusion on the surface (Fig. $\ref{VacAdatom11}$) and
sub-surface vacancy formation ( Fig. $\ref{VacAdatom52}$ and
$\ref{VacAdatom13}$). Fig. $\ref{VacAdatom345}$ corresponds to ad-atom
migration by exchange process. Some of the transition events are
affected by the location of ad-atom and vacancy. For example, as shown
in Fig. $\ref{VacAdatom157}$ and Fig. $\ref{VacAdatom601}$, the
formation of a divacancy and a nearest-neighbor ad-atom pair has lower
barrier than formation of a divacancy and two ad-atoms. Fig. $\ref{VacAdatom100}$
corresponds to formation of a bulk vacancy and an ad-atom.

\begin{figure}[thbp]
   \centering
   \includegraphics[width=0.4\textheight]{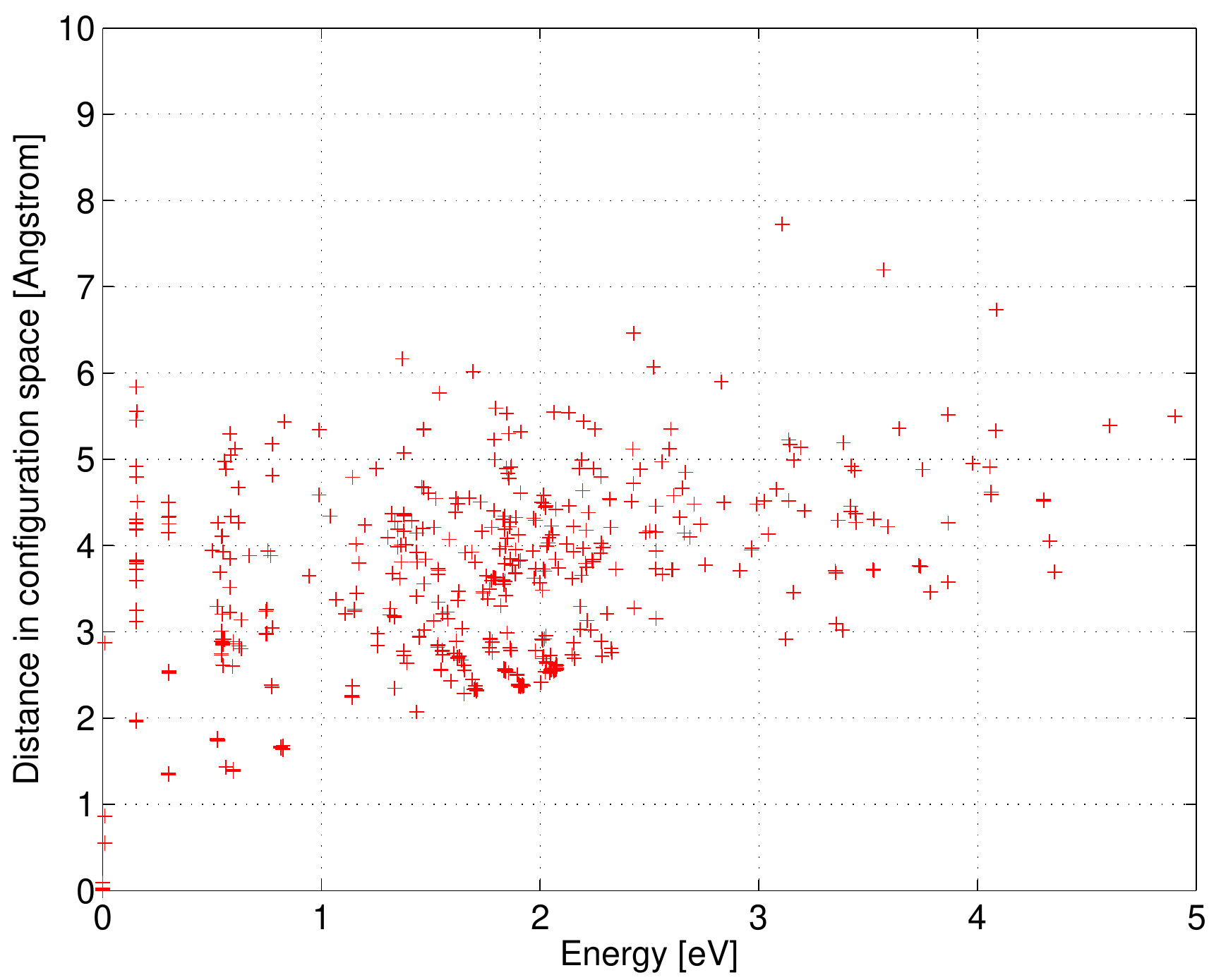}
   \caption{Summary of results depicting a wide spectrum of converged energy
     barriers for different rare events as a function of the distance
     of the final saddle configuration from the initial locally stable
     configuration for the surface vacancy and ad-atom system. The
     results correspond to simulations performed with $q_{c} = 0.50$
     thus allowing the possibility of sampling a much larger portion
     of the configuration space for index-1 saddle points.} 
   \label{DOStotVacAdatom}
 \end{figure}

Fig. $\ref{DOStotVacAdatom}$ shows a summary of the results obtained
from converged GAD simulations. The results are depicted as the
distribution of the converged saddle energies as a function of
distance of the converged saddle configuration from the initial
configuration in the configuration space. As can be seen from the
figure, there are multiple saddle configurations with same energy but
separated by a finite distance in the configuration space. 
%One such
%example for ad-atom diffusion case is when it converges at different
%saddle points as it hops between fcc and hcp sites on the $(111)$
%surface. In this case the system passes through many fixed points on
%the PES. 

\begin{figure}[thbp]
   \centering
   \subfigure[] {
     \includegraphics[width=0.4\textheight]{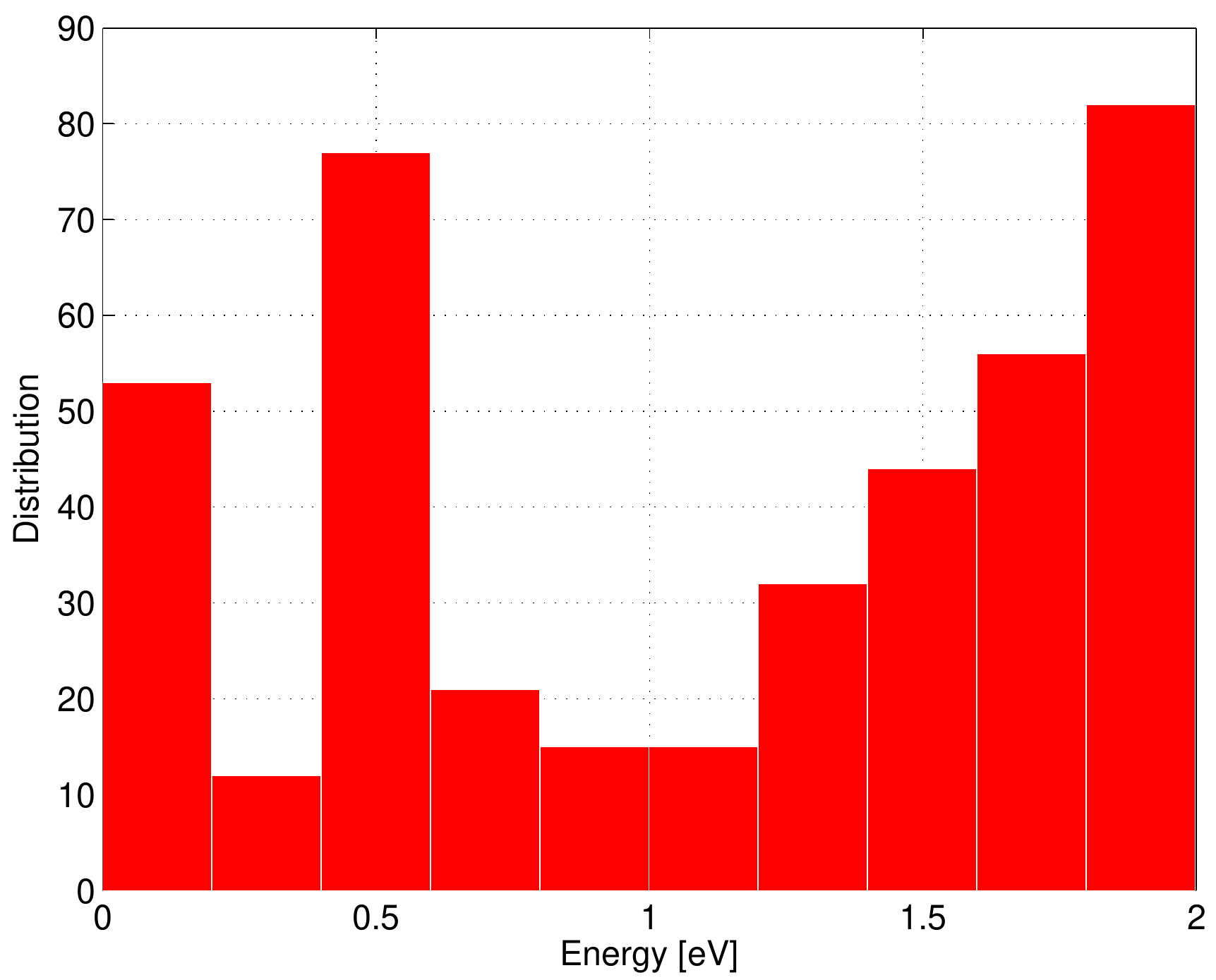}
     \label{DOStotVacAdatom40}
   }
   \subfigure[]  {
     \includegraphics[width=0.4\textheight]{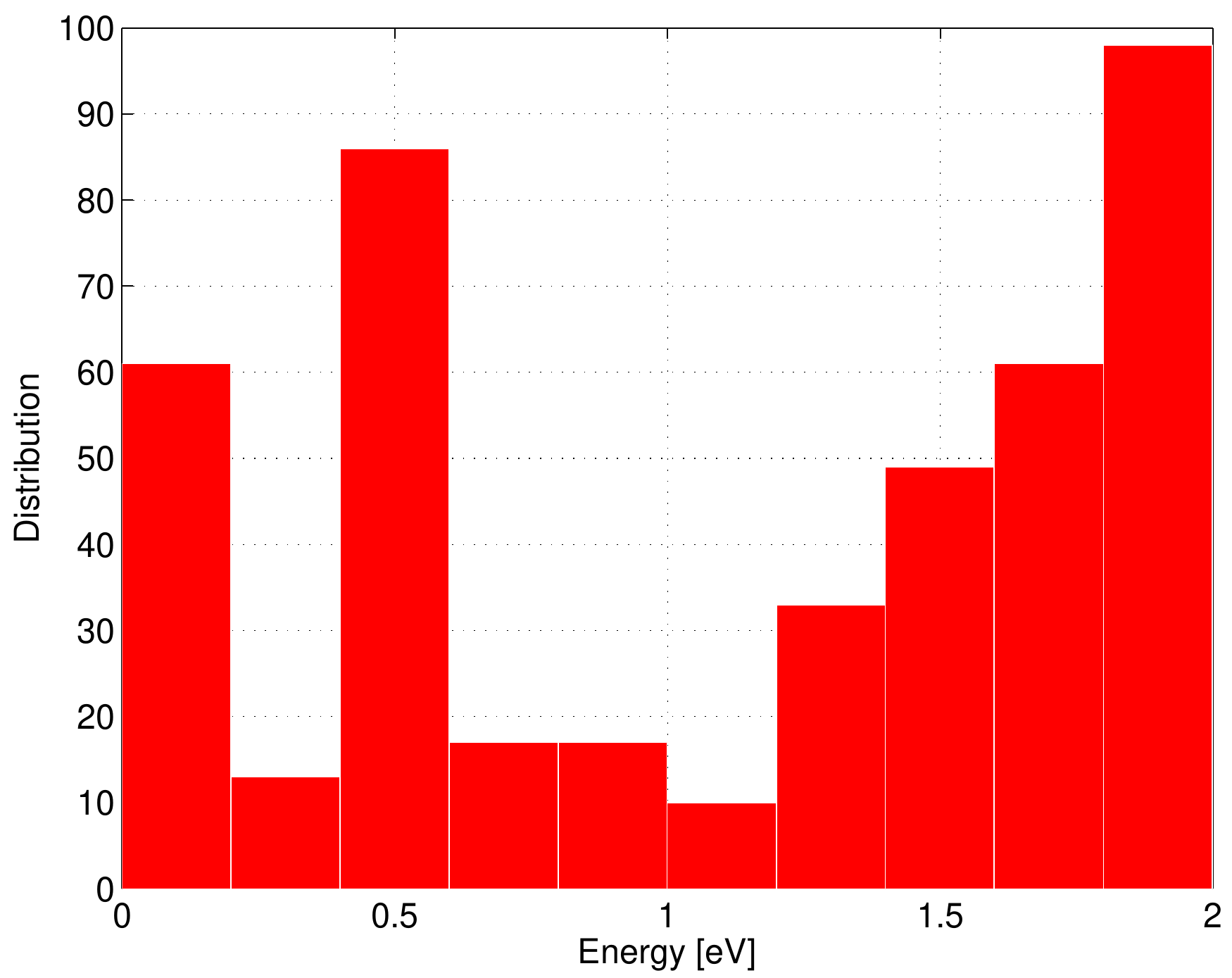}
     \label{DOStotVacAdatom30}
   }
   \caption{Summary of results depicting the distribution of converged
     saddle energies (with respect to the initial structure) for $q_{c} = 0.40$
     ($\ref{DOStotVacAdatom40}$) and $q_{c} = 0.30$
     ($\ref{DOStotVacAdatom30}$) for the system with a surface vacancy
     and ad-atom. The results correspond to the evolution equations ($\ref{EqMotion1}$)}
   \label{DOSdistQc}
 \end{figure}

Fig. $\ref{DOSdistQc}$ shows the distribution of converged saddle
energies less than 2.0 eV for different values of $q_{c}$ for a fixed
set of initial direction vectors. This suggests that, for a
sufficiently large set of well sampled direction vectors, the values
of $q_{c}$ in the range of 0.30-0.50 provides a good collection of low
barrier saddle points. Though our selection of initial direction
vectors significantly decreases repeated visits of the same saddle point,
however on a high dimensional PES there exists saddle points with
similar barriers but separated by a finite distance. The above results
have significant contributions from such scenarios. 

\begin{figure}[thbp]
   \centering
   \subfigure[]  {
     \includegraphics[width=0.225\textheight]{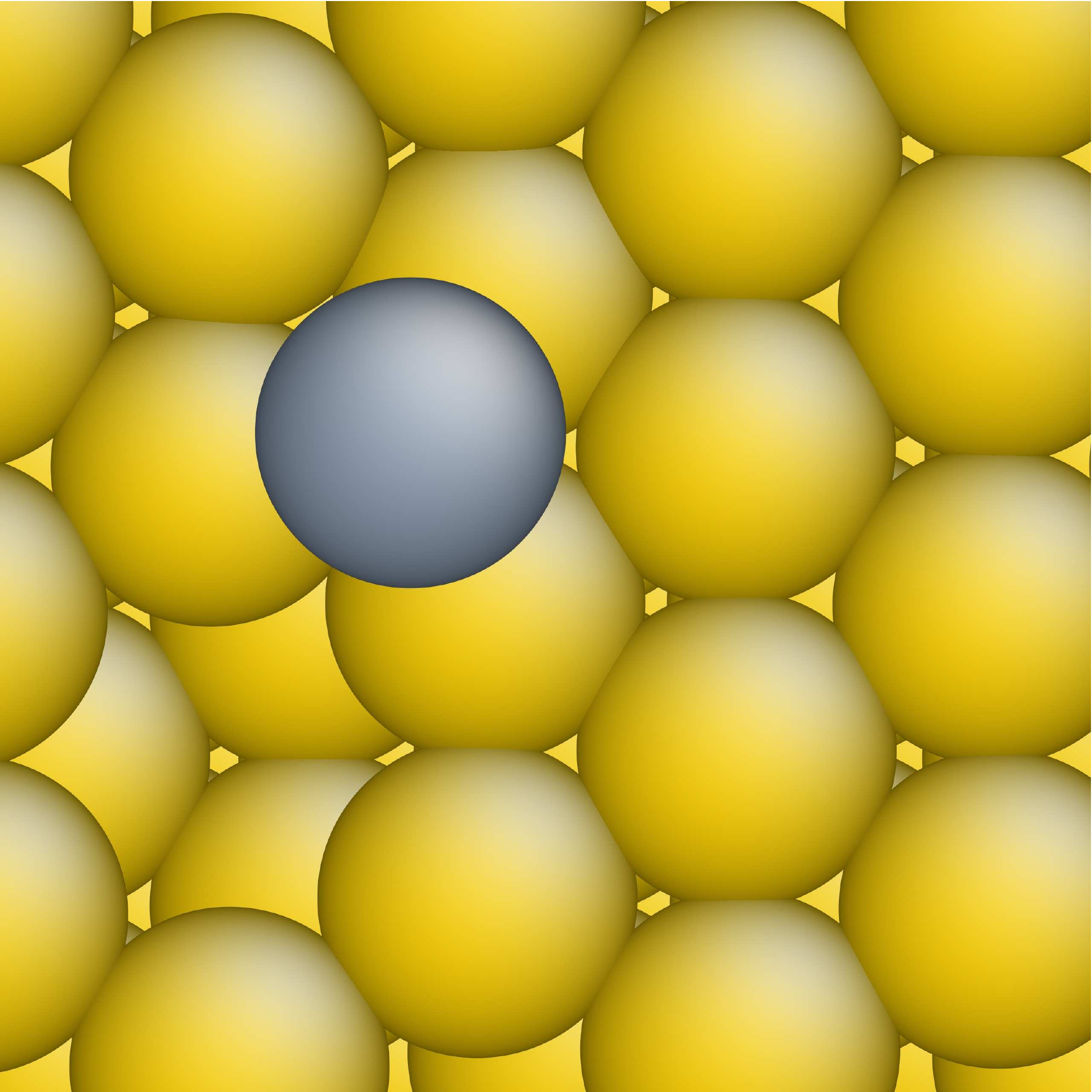}
     \label{AdatomMigration1}
   }
   \subfigure[] {
     \includegraphics[width=0.225\textheight]{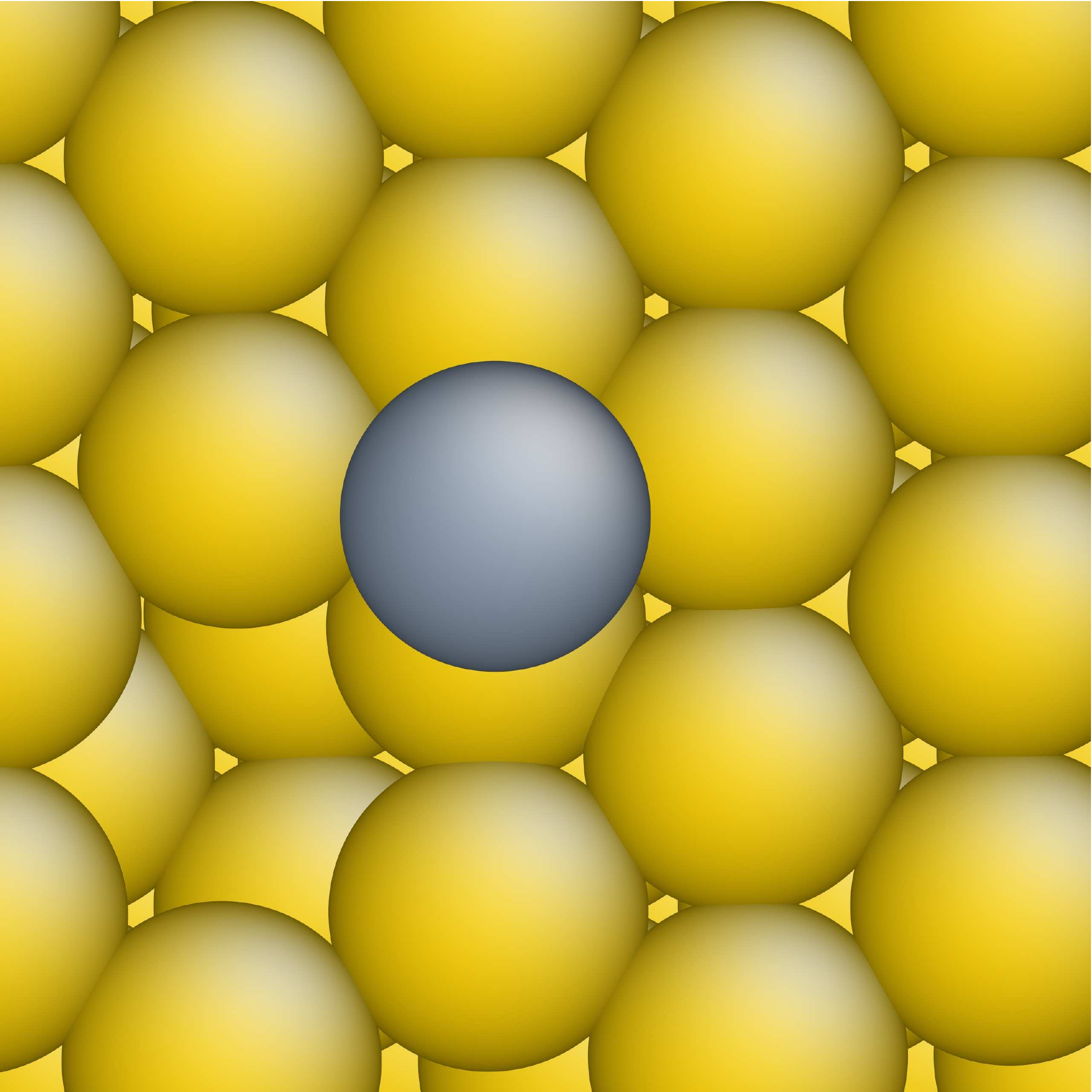}
     \label{AdatomMigration2}
   }
   \subfigure[] {
     \includegraphics[width=0.225\textheight]{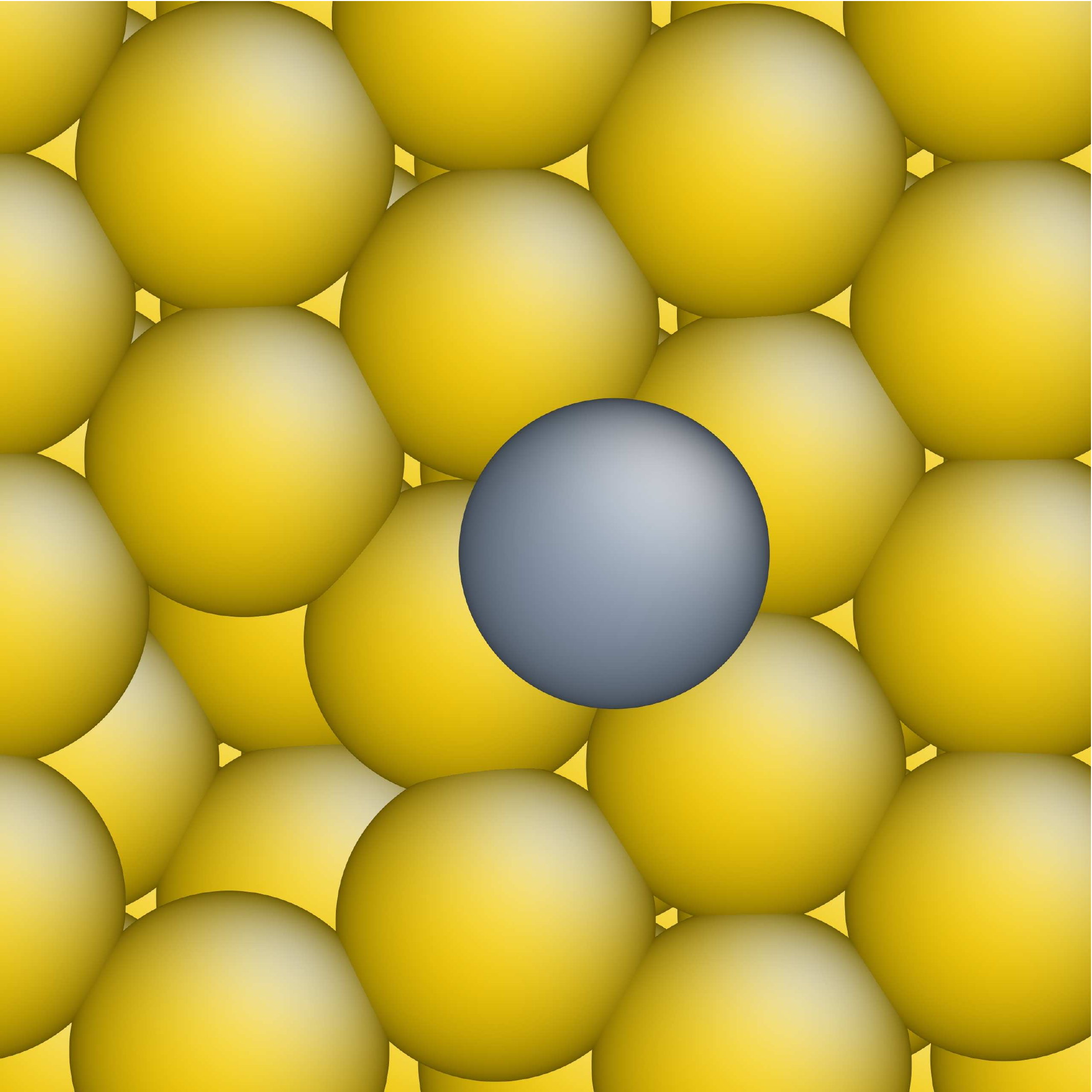}
     \label{AdatomMigration3}
   }
   \subfigure[] {
     \includegraphics[width=0.225\textheight]{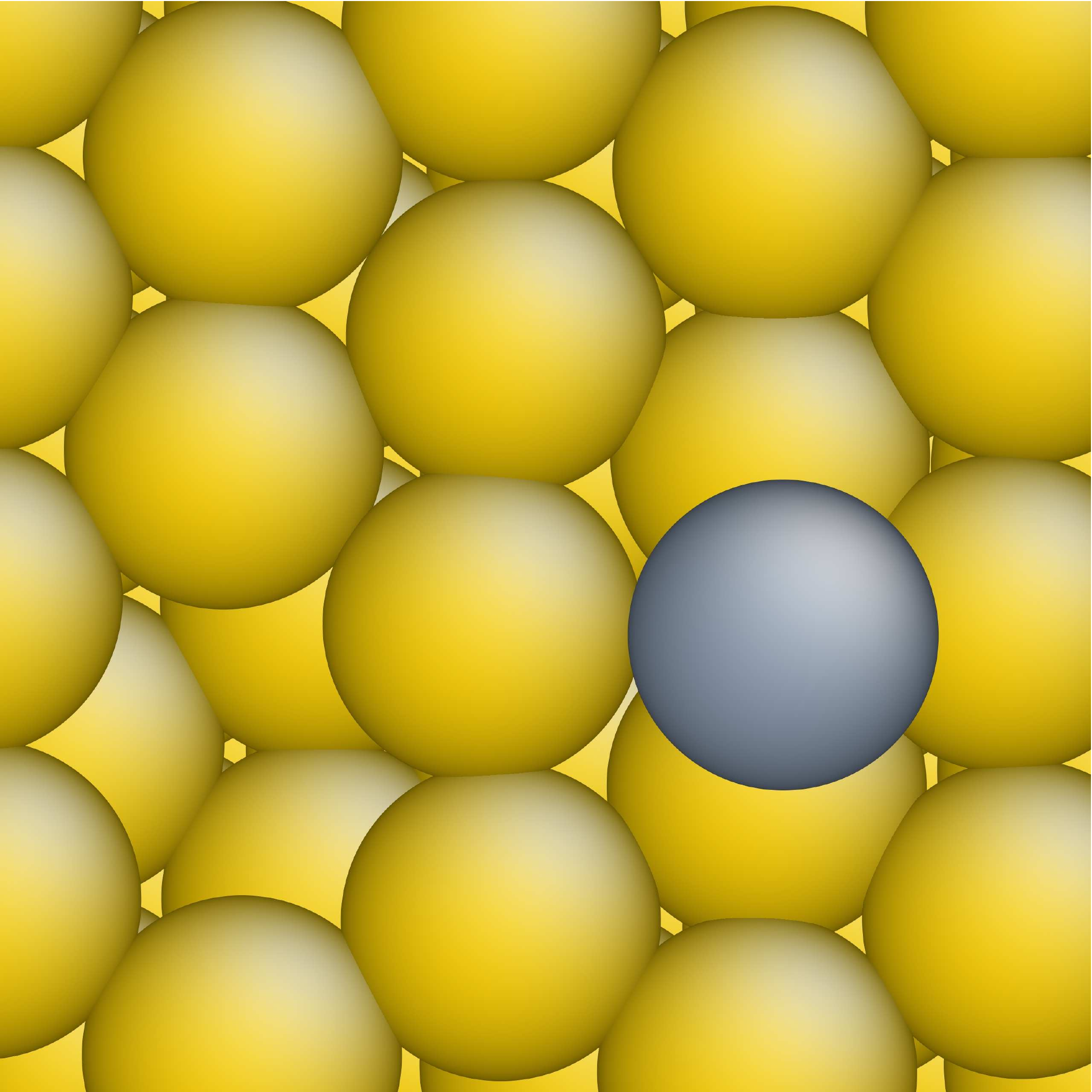}
     \label{AdatomMigration4}
   }
   \caption{Sampling of saddle points using MD-GAD. The adatom (shown
     in grey) in this case hops between different index-1 saddle
     points (Fig. $\ref{AdatomMigration2}-\ref{AdatomMigration4}$) on
     the $\langle 111\rangle$ surface of
     Cu. Fig. $\ref{AdatomMigration1}$ is the initial structure.} 
   \label{MDgadAdatomVac}
 \end{figure}

Fig. $\ref{MDgadAdatomVac}$ shows a collection of saddle points
obtained by using MD-GAD. In this case, the simulations are started
with zero initial velocity from the local minimum in
Fig. $\ref{AdatomMigration1}$ and the direction vector preferentially
initialized to activate the ad-atom. As the simulation progresses, the
system moves from the local minimum point to a saddle point
(Fig. $\ref{AdatomMigration2}$) where the KE reaches a maximum
value. Then the high KE pushes the system to the next saddle point
shown in Fig. $\ref{AdatomMigration3}$. Thus MD-GAD effectively
samples many low lying saddle points on the PES.

\subsection{Heptamer island on (111) surface of copper}

The simulation setup consists of a sample with 487 Cu atoms and lattice
vectors parallel to $[110]$, $[1\bar{1}1]$ and $[\bar{1}12]$
directions. The periodic boundary conditions are imposed along $[110]$ and
$[\bar{1}12]$ while free surface conditions are maintained along $[1\bar{1}1]$
direction. The atomic interactions are modeled using an EAM potential
developed by Mishin et al.$\cite{MishinFMP99}$ Initially all atoms in
the seven atom heptamer island occupy fcc sites on the $(111)$ surface.

\begin{figure}[thbp]
   \centering
   \subfigure[] {
     \includegraphics[width=0.24\textheight]{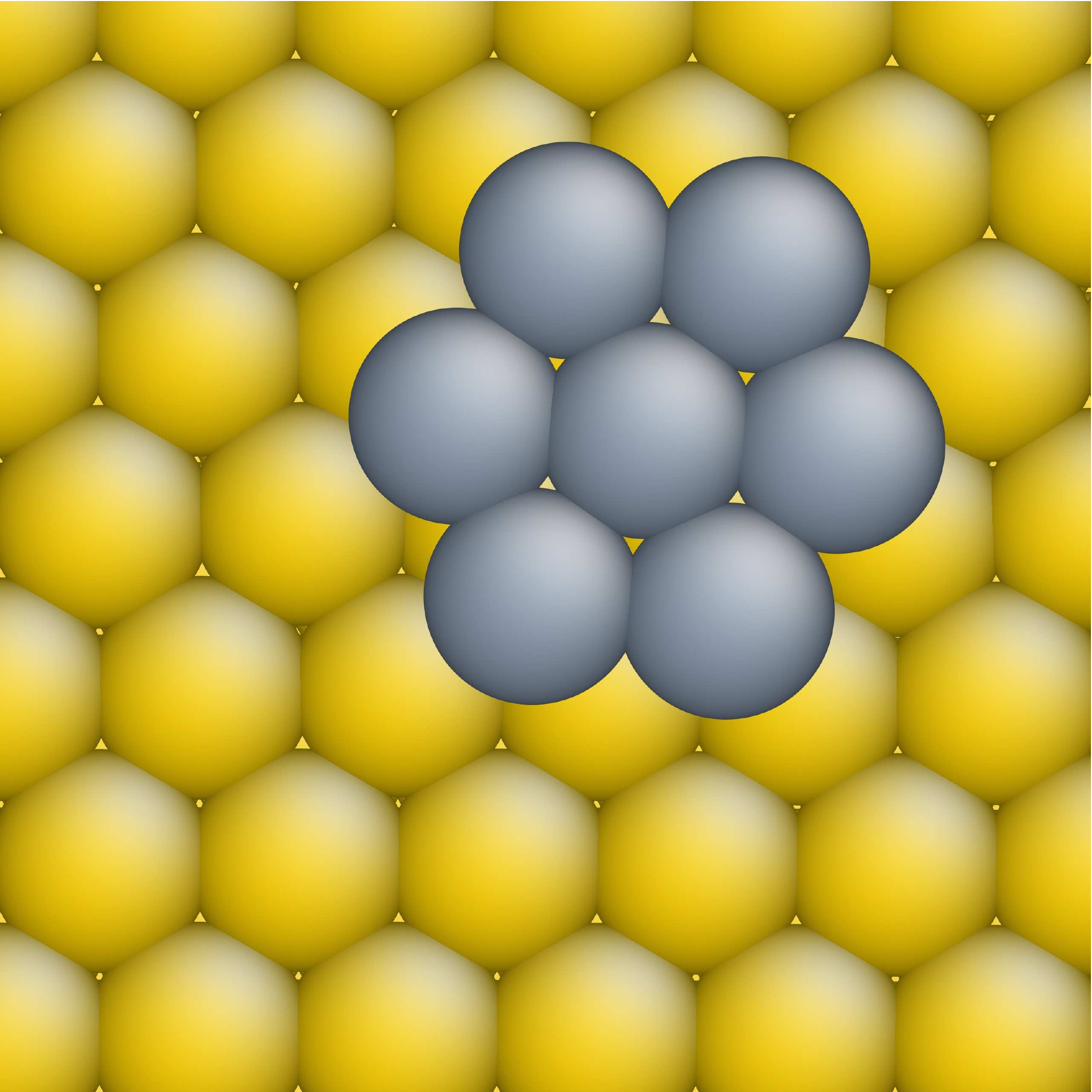}
     \label{Heptamer743a}
   }
   \subfigure[] {
     \includegraphics[width=0.24\textheight]{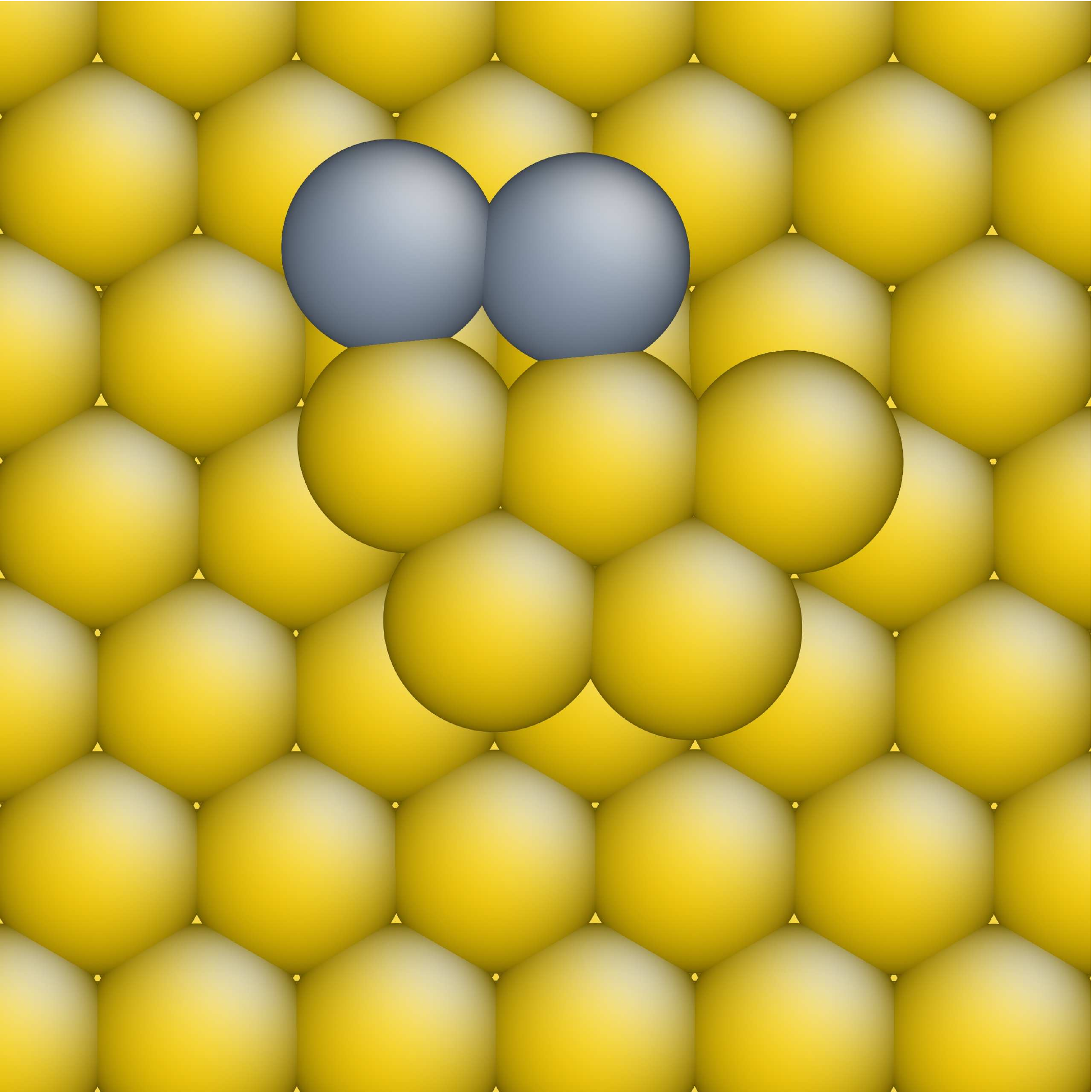}
     \label{Heptamer75}
   }
   \subfigure[] {
     \includegraphics[width=0.24\textheight]{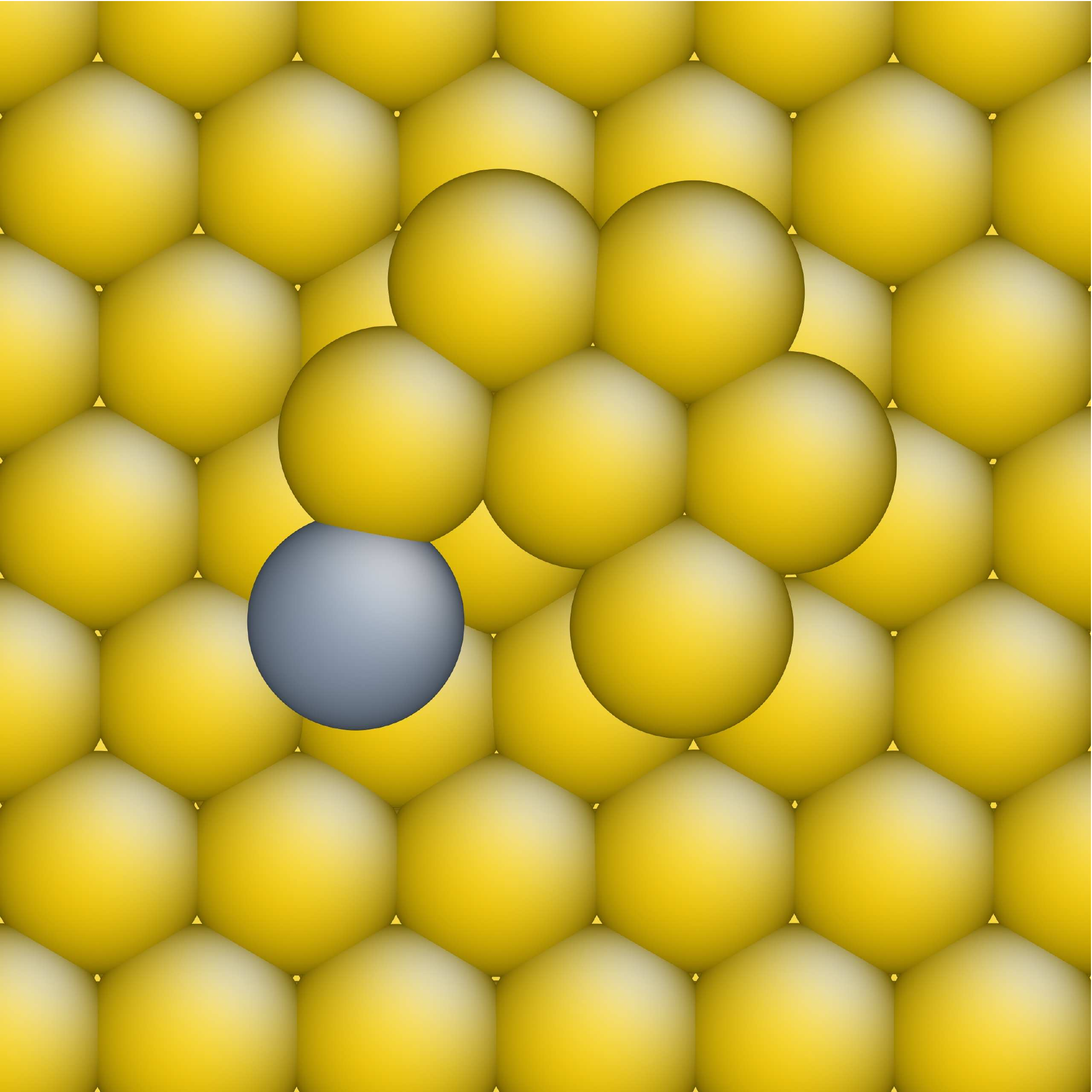}
     \label{Heptamer0}
   }   
   \subfigure[] {
     \includegraphics[width=0.24\textheight]{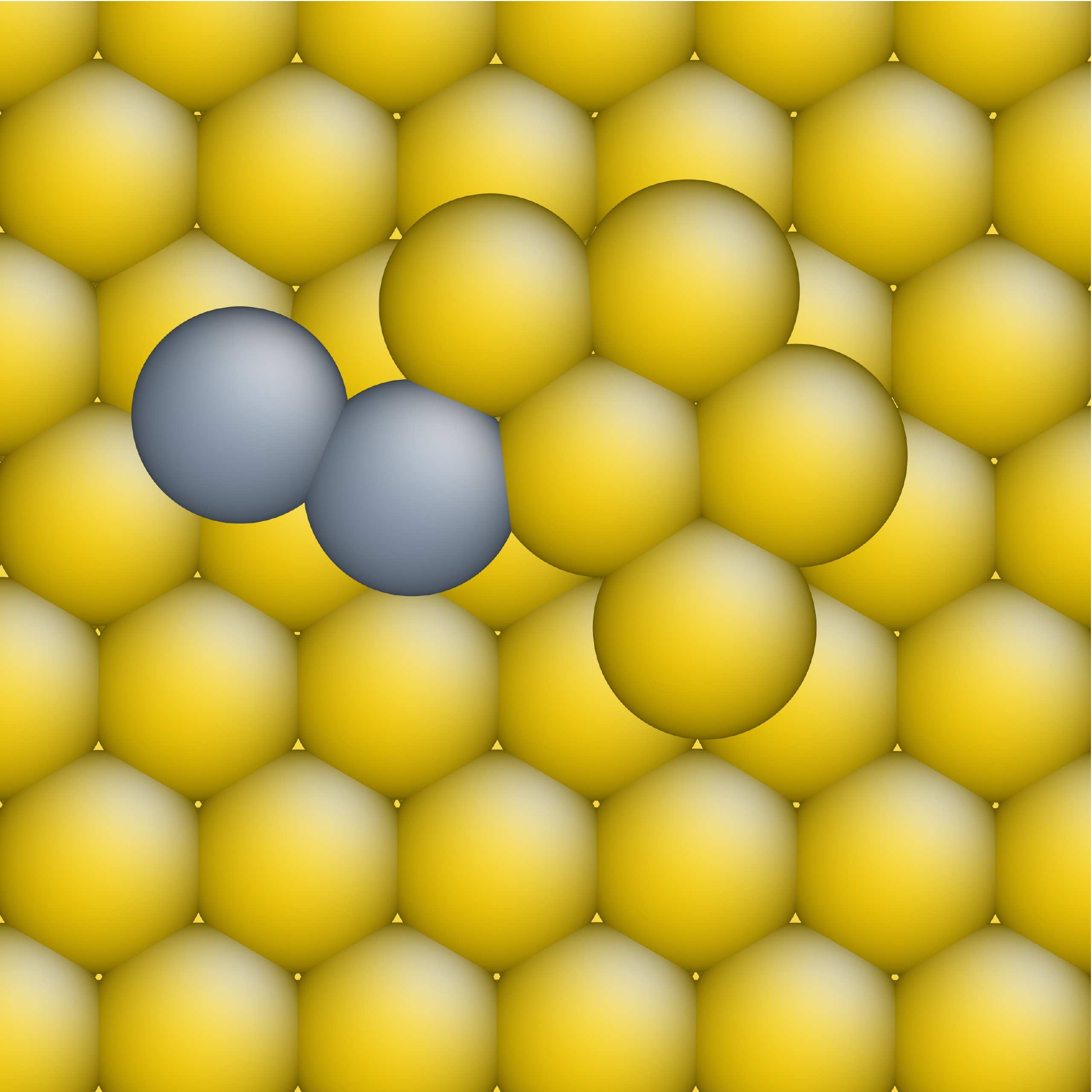}
     \label{Heptamer2}
   }   
   \subfigure[] {
     \includegraphics[width=0.24\textheight]{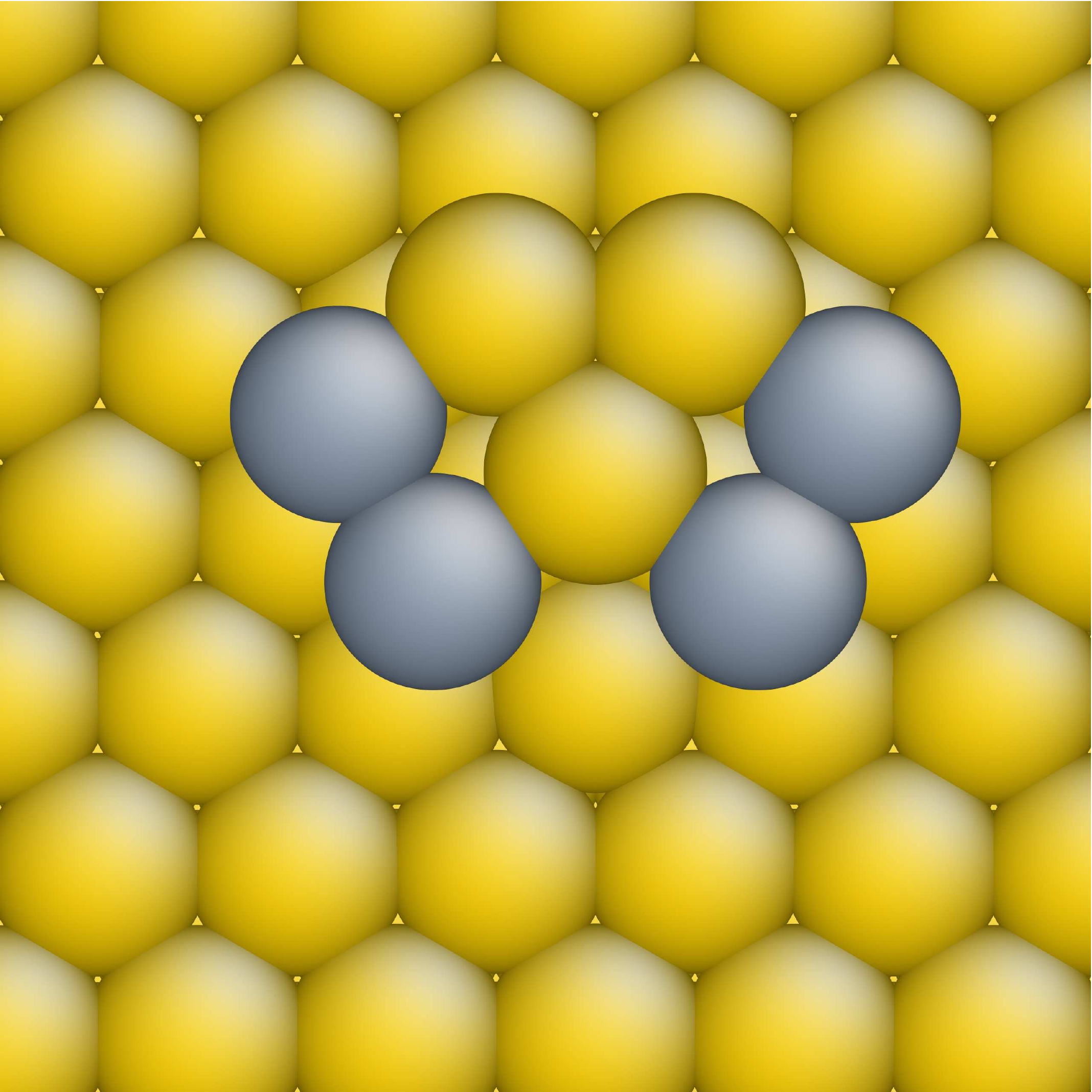}
     \label{Heptamer98}
   }
   \subfigure[] {
     \includegraphics[width=0.24\textheight]{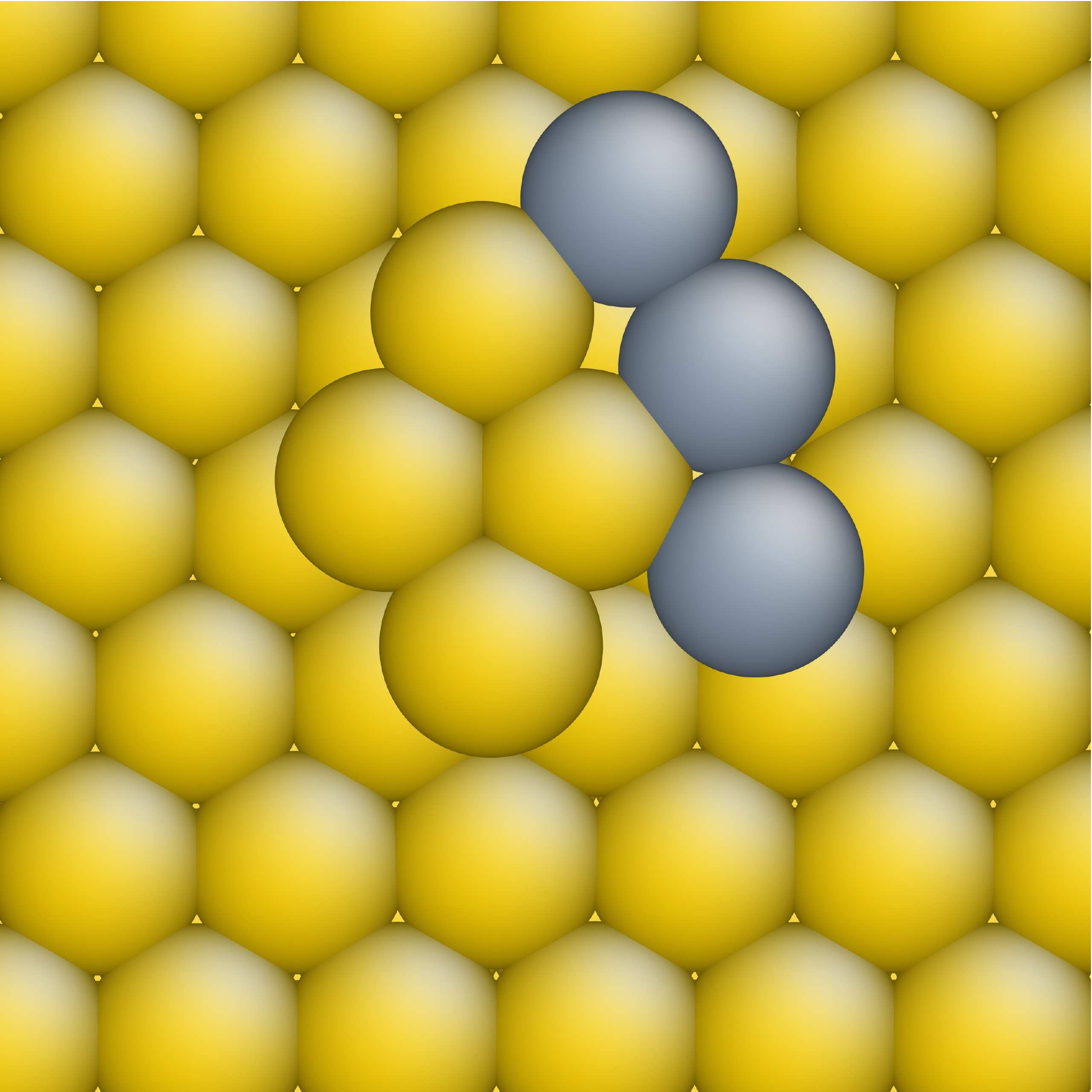}
      \label{Heptamer638}
   }
   \caption{Some converged saddle configurations for the vacancy ad-atom
     configuration obtained from GAD simulations. $\ref{Heptamer743a}$ island
     sliding on (111) surface (barrier 0.39 eV), $\ref{Heptamer75}$
     two atom sliding (barrier 0.49 eV), $\ref{Heptamer0}$ single atom
     hopping (barrier 0.683), $\ref{Heptamer2}$ two atom sliding
     (barrier 0.69 eV), $\ref{Heptamer98}$ two atom sliding process
     taking place at two locations on the island (barrier 0.78 eV),
     $\ref{Heptamer638}$ sliding of atoms (barrier 0.84 eV).} 
 \end{figure}

\begin{figure}[thbp]
   \centering
   \subfigure[]  {
     \includegraphics[width=0.24\textheight]{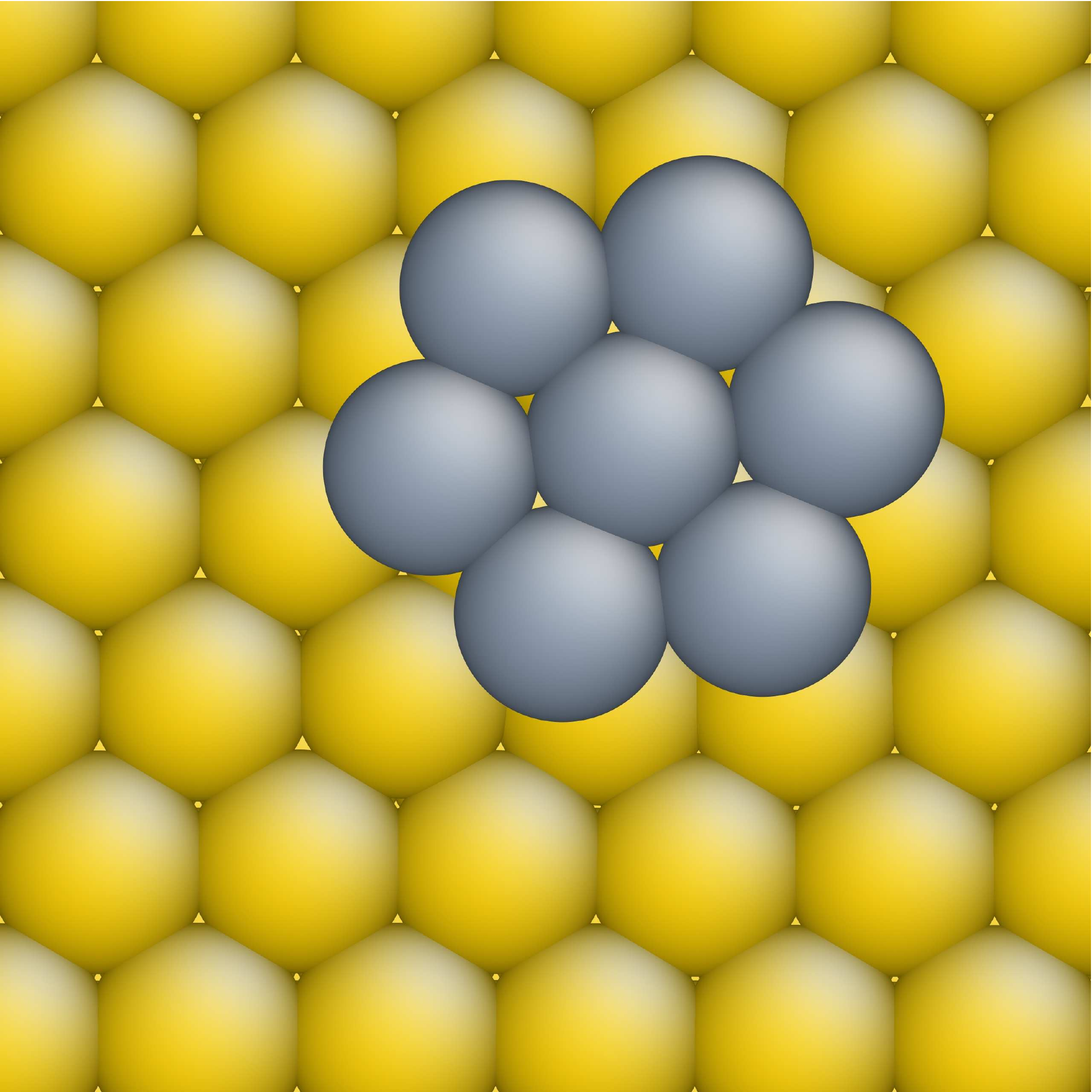}
     \label{Heptamer742}
   }
   \subfigure[] {
     \includegraphics[width=0.24\textheight]{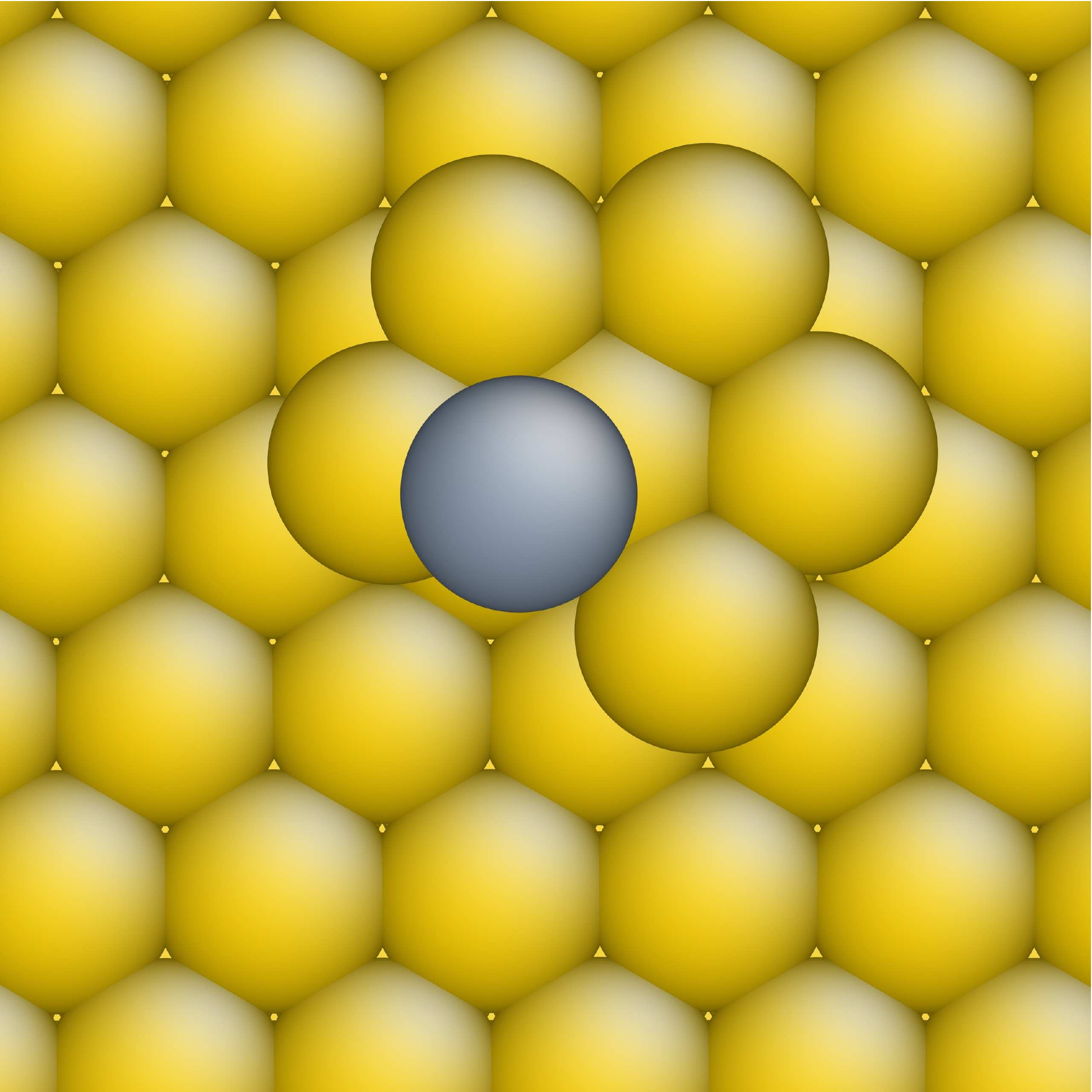}
     \label{Heptamer1}
   }
   \subfigure[] {
     \includegraphics[width=0.24\textheight]{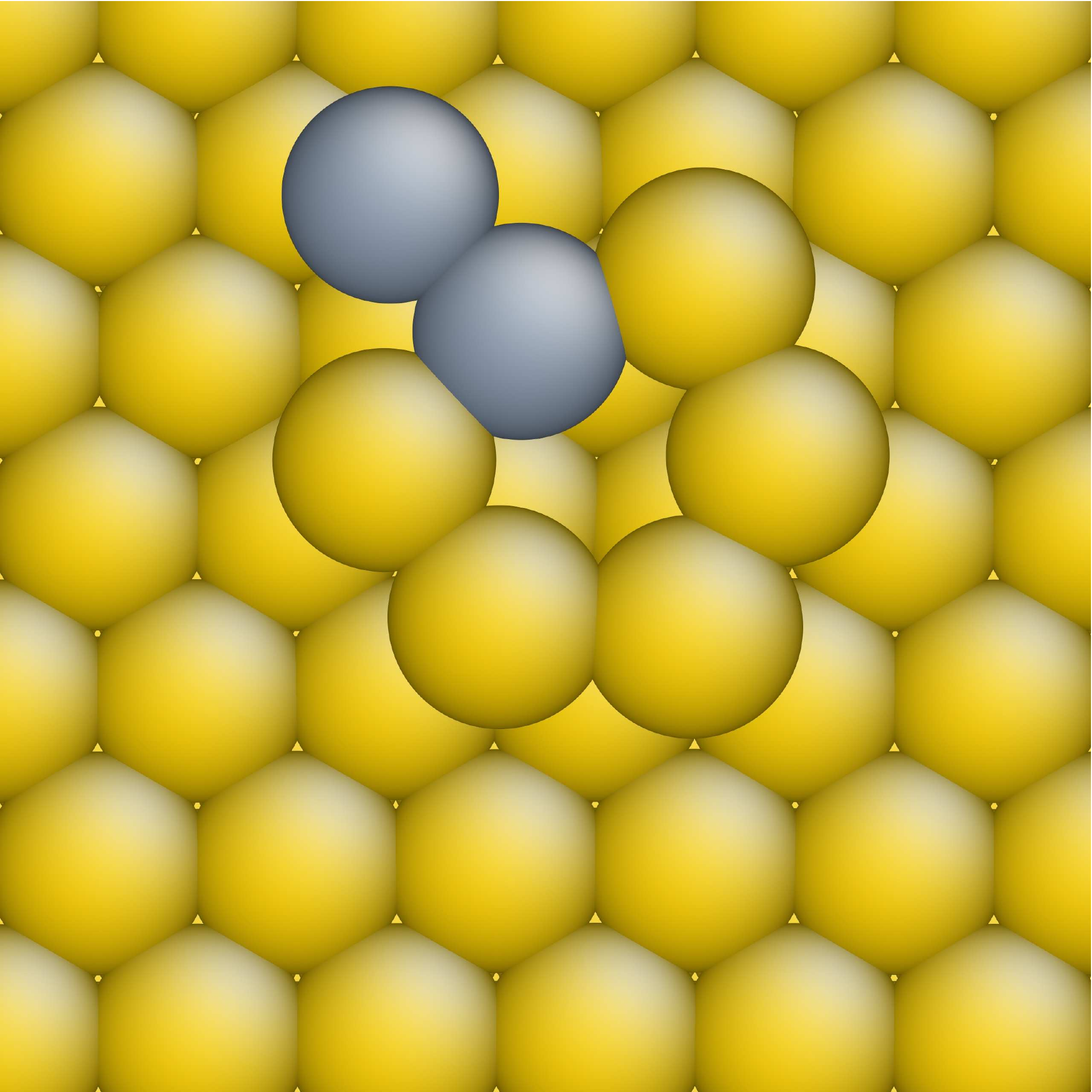}
     \label{Heptamer378}
   } 
   \subfigure[] {
     \includegraphics[width=0.24\textheight]{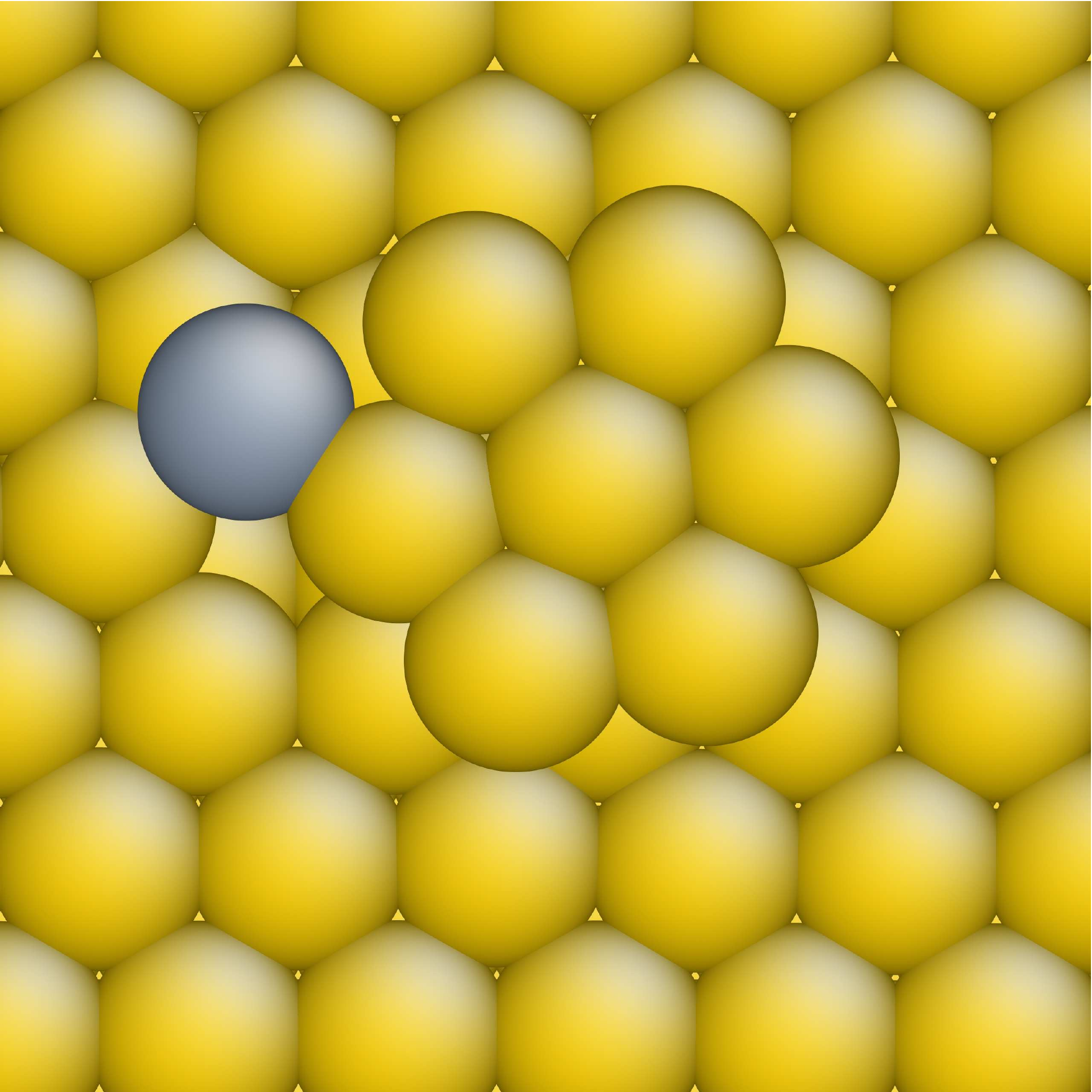}
     \label{Heptamer20}
   }
   \caption{Some converged saddle configurations for the vacancy ad-atom
     configuration obtained from GAD simulations. $\ref{Heptamer742}$
     island rotation (barrier 0.92 eV), $\ref{Heptamer1}$ ad-atom formation on
     heptamer island (barrier 1.39 eV), $\ref{Heptamer378}$ two atom
     sliding process (barrier 1.49 eV), $\ref{Heptamer20}$
     ad-atom and surface vacancy formation (barrier 1.86 eV).}   
 \end{figure}

\begin{figure}[thbp]
   \centering
   \subfigure[]  {
     \includegraphics[width=0.24\textheight]{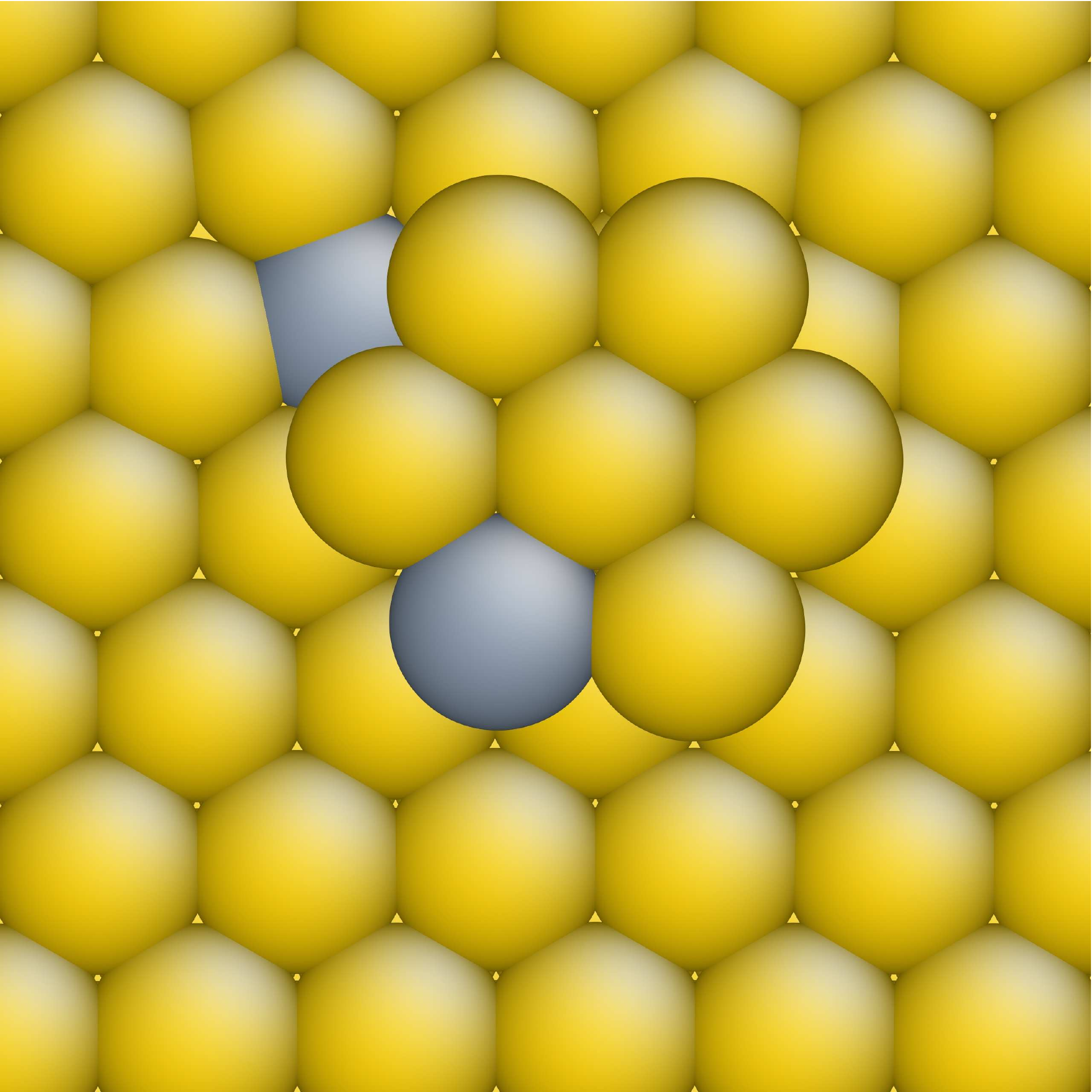}
     \label{Heptamer23a}
   }
   \subfigure[] {
     \includegraphics[width=0.24\textheight]{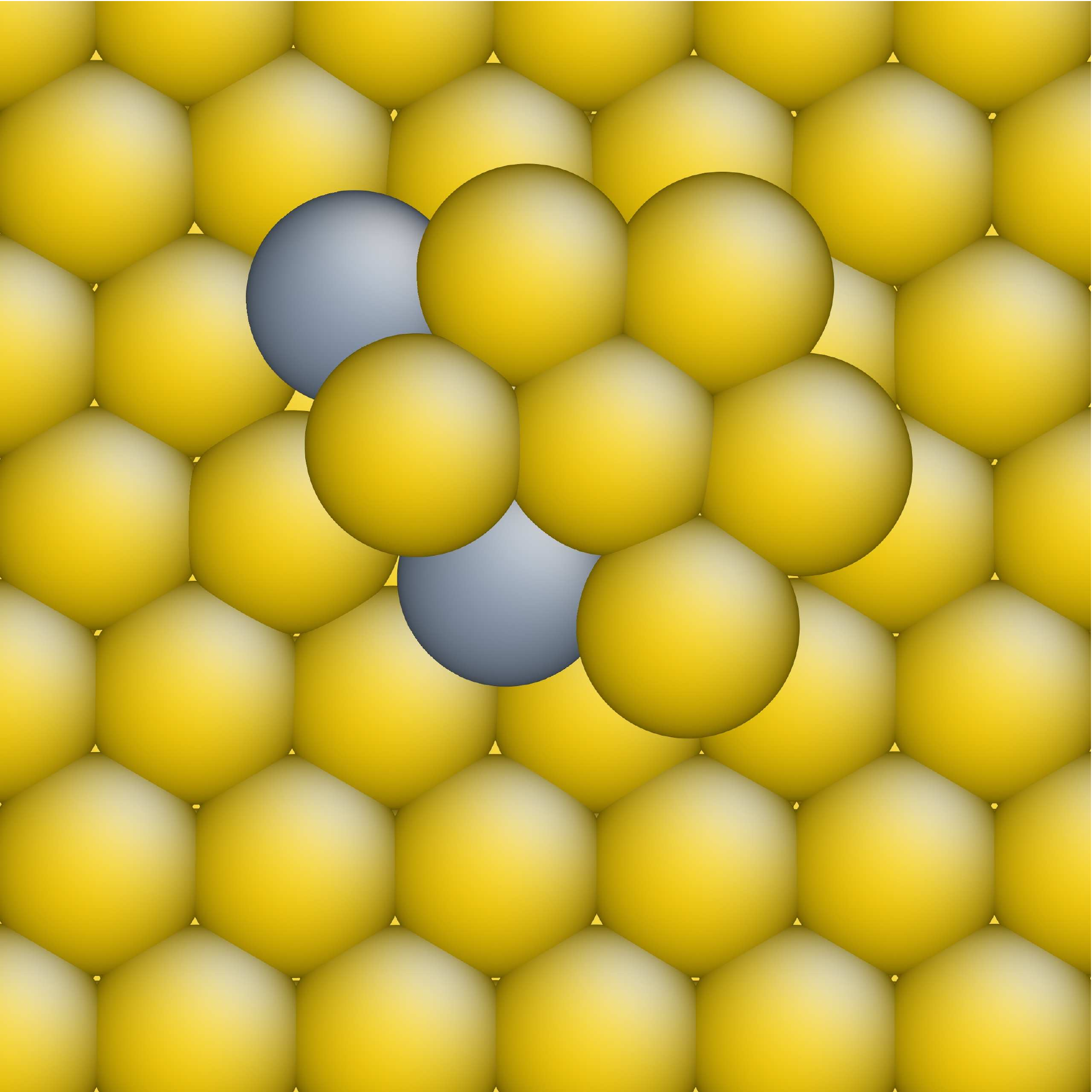}
      \label{Heptamer23b}
   }
   \caption{$\ref{Heptamer23a}$ ad-atom formation by exchange
     mechanism which changes the shape of the heptamer island
     (barrier 1.84 eV).}
   \label{Heptamer23}
 \end{figure}

Fig. $\ref{Heptamer743a}$ to Fig. $\ref{Heptamer20}$ shows a collection
of, low barrier, saddle configurations. From our simulations, we find
many collective processes, such as those involving lateral translation
(energy barrier 0.39 eV) and rotation (energy barrier 0.92 eV) of the
heptamer island. These collective processes have also been reported for
islands of different sizes and shapes.$\cite{MironF01}$ 
Other low barrier events include sliding of two nearest neighbor atoms
belonging to the heptamer island (Figs. $\ref{Heptamer75}$,
$\ref{Heptamer2}$, $\ref{Heptamer98}$ and $\ref{Heptamer378}$) all
with barriers 0.49-0.69 eV. Fig. $\ref{Heptamer638}$ involves a
cooperative rearrangement and sliding of three
atoms and has little higher barrier. The sliding of group of atoms with respect to others in the
island resembles the process of formation of a
dislocation in a bulk crystal. We also observe movement of an island
by repeated shearing of different layers.$\cite{MironF01}$ Fig. $\ref{Heptamer23}$ and
$\ref{Heptamer20}$ shows different ways of formation of a surface vacancy and an
ad-atom. While the process in Fig. $\ref{Heptamer20}$ involves a surface atom hopping
out of plane to attach itself to the island the process in Fig.$\ref{Heptamer23}$ involves a
cooperative ``exchange'' mechanism involving a sub-surface atom.$\cite{PedersenHJ11}$

\begin{figure}[thbp]
   \centering
   \includegraphics[width=0.4\textheight]{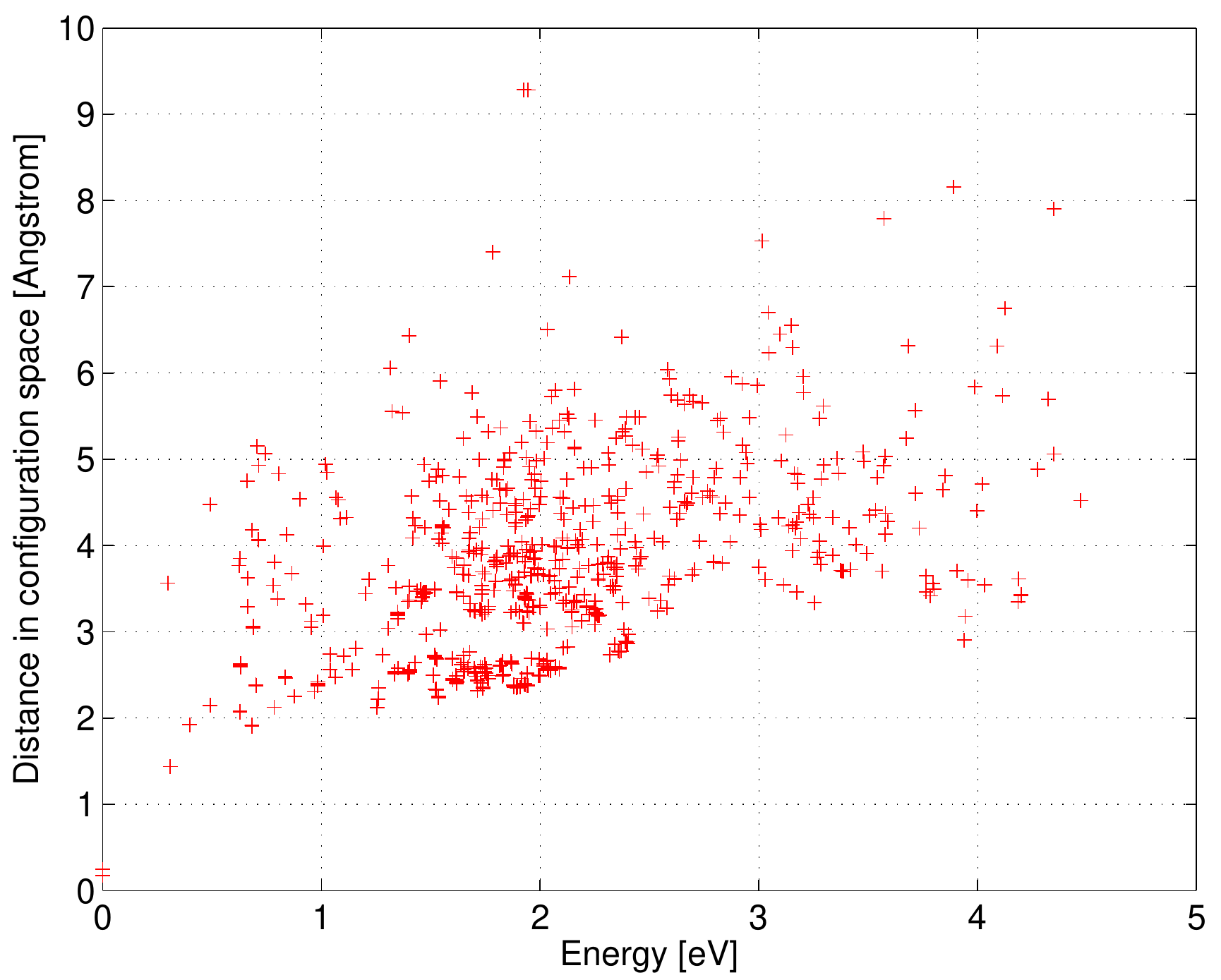}
   \caption{Summary of results depicting a wide spectrum of converged energy
     barriers for different rare events as a function of the distance
     of the final saddle configuration from the initial locally stable
     configuration for the system with a heptamer island on $(111)$ surface.}
   \label{DOStotalHeptamer}
 \end{figure}

\begin{figure}[thbp]
   \centering
   \subfigure[]  {
     \includegraphics[width=0.24\textheight]{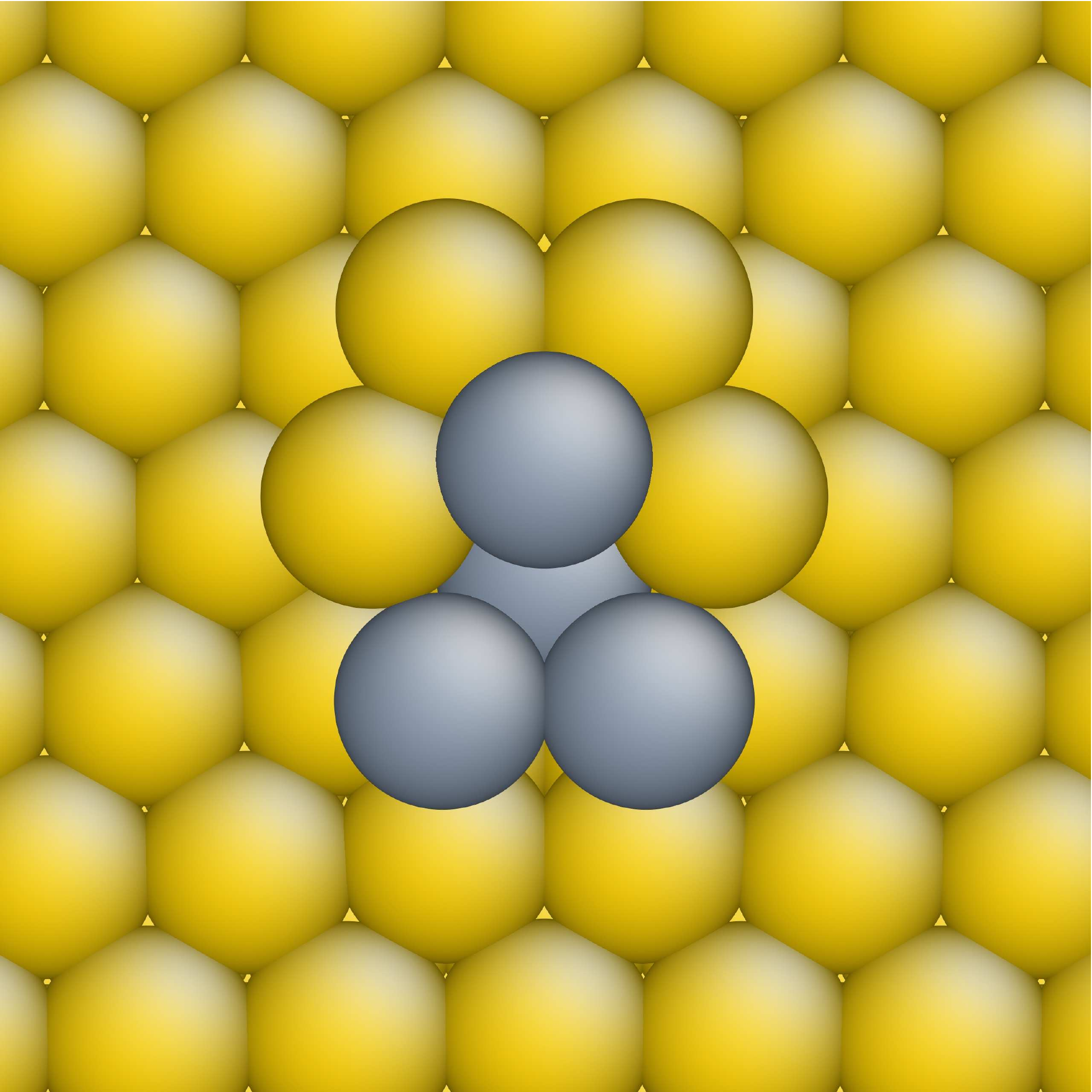}
     \label{Heptamer50a}
   }
   \subfigure[] {
     \includegraphics[width=0.24\textheight]{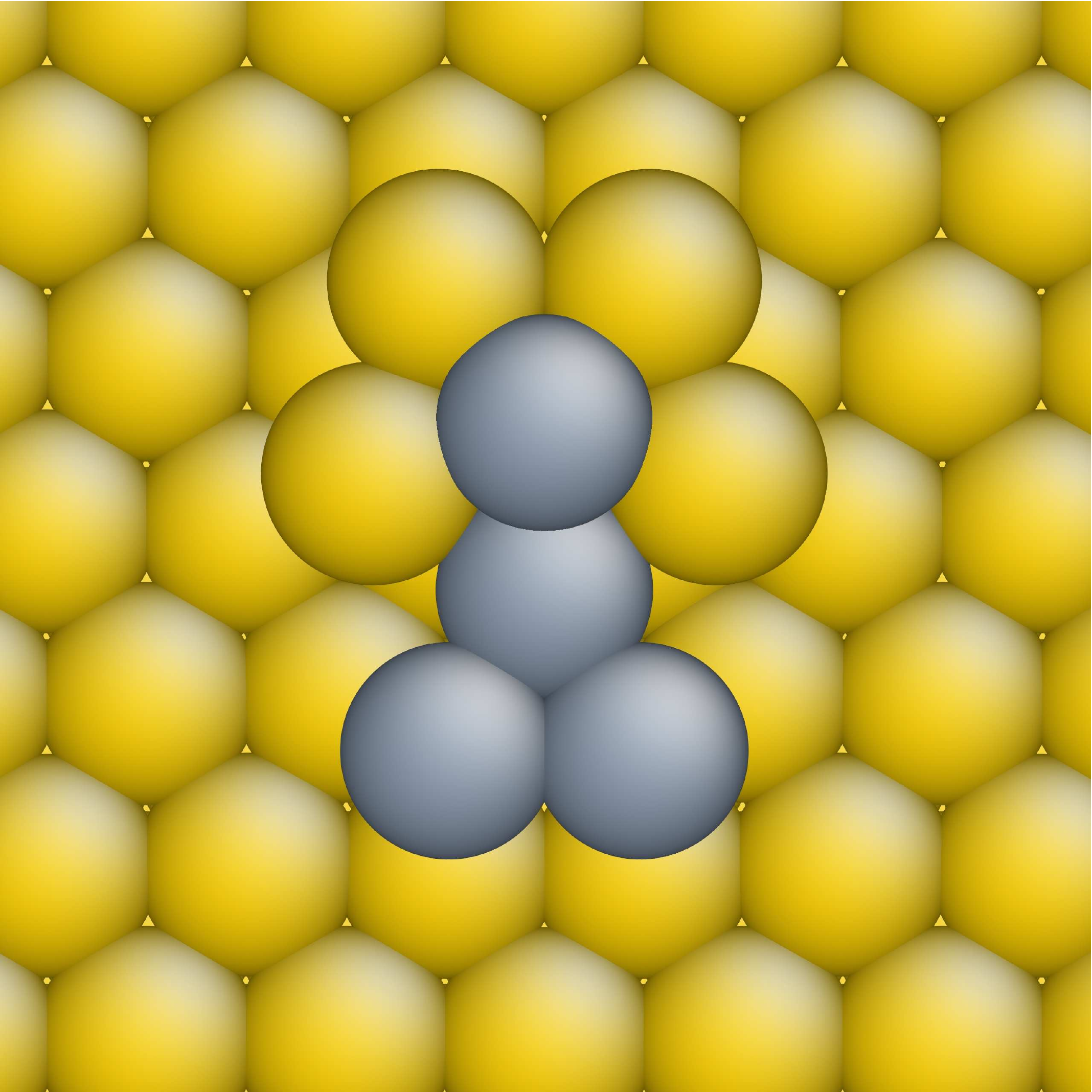}
      \label{Heptamer50b}
   }
   \caption{Snapshots of the collective process of formation of a sub-surface
     vacancy and an ad-atom on top of the heptamer island. The atoms
     involved in the process are shown in grey (barrier 2.42 eV).}
   \label{Heptamer50}
 \end{figure}

Fig. $\ref{DOStotalHeptamer}$ shows a summary of the converged
saddle configuration energies, obtained using the 843 initial
direction vectors, as a function of the distance in the
configuration space from the initial local minimum configuration. 
Setting the parameter $q_{c}=0.50$ provides the flexibility to sample
a larger portion of the configuration space near the initial
configuration. This also allows us to capture events which involve
multiple saddle points thus providing vital information about possible
channels for probable transition events. Fig. $\ref{Heptamer50}$
shows one such example where a sequence of transition events leads to
the formation of a bulk vacancy, an ad-atom on top of the heptamer
island and subsequent change in shape of the island on the $(111)$
surface.

\section{Conclusions}

We have shown that the deterministic, molecular dynamics or stochastic
versions of GAD can be used as effective tools for sampling saddle
points on the PES. The algorithms presented here are scalable to
systems of higher dimensions. The computational cost and memory
requirements are similar to molecular dynamics simulations. 

A method to initialize the direction vectors along which the
system can move out of the initial potential energy well is described.
It is based on the knowledge of the vulnerable DOF of the system from
the force at initial local minimum. All the operations in GAD and its
finite temperature variants have $\mathcal{O}\left(N\right)$ ($N$ is
the number of DOF) computational complexity. Our use of a criteria based
on equilibrium bond length of atoms in a system to identify the closeness 
of a system from the initial local minimum point also saves
computational cost.

The finite temperature versions of GAD open avenues for sampling of
saddle points on the PES. Generally existing macroscopic models are
based on information obtained from locally stable fixed points of a
system. In the stochastic version of GAD, since the system can hop
between different low energy saddle points, a wealth of information
can be obtain about the saddle configurations. This information can
then be used to improve the existing mean field models.

GAD can be extended to handle higher index saddle points, as was done
in Ref[15]. It is obvious that one can also extend the work presented
to those cases. 

Our results for point defect activity on a $(111)$ surface
obtained from GAD show a variety of processes involving more than a
single atom. Indeed the wide spectrum of diffusion processes revealed
by GAD for ad-atoms, vacancies and islands can provide much needed 
insight into the probable mechanisms behind many experimentally
observed surface phenomena.$\cite{TsongC92, FerronGdM04,
  vanGastelSvvF01, MontalentiVF02}$

\section{Appendix I:Local convergence analysis for MD-GAD}

The fixed points of PES are also the fixed points
of ($\ref{EqMotion4}$). To see this consider a two dimensional
system with force ${\bf F} = f_{x}\hat{i} + f_{y}\hat{j}$ and
direction vector ${\bf n} = n_{x}\hat{i} +  n_{y}\hat{j}$, where
$f_{x}$, $f_{y}$ and $n_{x}$, $n_{y}$ are individual components of
force and direction vectors, respectively, along the $i$ and $j$
axis. Consequently, the solutions of $\dot{\bf v} = 0$ are: 
\begin{equation}
  f_{x}\hat{i} + f_{y}\hat{j} = 2\lambda\left(n_{x}\hat{i} +
    n_{y}\hat{j}\right),\qquad\lambda = f_{x}n_{x} + f_{y}n_{y}\;,\qquad
  n_{x}^{2} + n_{y}^{2} = 1
\end{equation}
Equating the individual vector components on both sides of the above
equation we obtain
\begin{equation}
  \left.\begin{array}{cc}
      f_{x}\left(n_{y}^{2}-n_{x}^{2}\right) & = 2f_{y}n_{y}n_{x}\\
      f_{y}\left(n_{x}^{2}-n_{y}^{2}\right) & = 2f_{x}n_{y}n_{x}\\
    \end{array}\;\right\}\qquad f_{x}^{2} + f_{y}^{2} = 0  
\end{equation}
which is possible only if individual components of the force vector
are zero. So for all values of the direction vector the above
condition is satisfied only at fixed points of $V\left(\bf x\right)$.

The Jacobian matrix of MD-GAD has the form :
\begin{equation}
  \tilde{\bf J}_{1} = \left[\begin{array}{ccc}
      0 & \mathbb{I} & 0\\
      \left(\bf J-2nn^{\rm T}J\right) & 0 & -2\left({\bf F^{\rm T}n}\mathbb{I} + 
        {\bf nF}^{\rm T}\right)\\
      \bf L & 0 & \left(\bf J- {2}nn^{\rm T}J - n^{\rm T}Jn\mathbb{I}\right) 
      \label{Dynamical2}
    \end{array}\right]
\end{equation}
where,$\;\bf L$ is a $n\times n$ matrix involving the third order
derivatives of $V$. At any fixed point of the potential, since the derivative of
$V$ is zero, the term ${\bf F^{\rm T}n}\mathbb{I} + {\bf nF}^{\rm T}$
is zero and the eigenvalues of $\tilde{\bf J}_{1}$ at these fixed
points can be obtained from 
\begin{equation}
  \tilde{\bf J}_{1} = \left[\begin{array}{ccc}
      0 & \mathbb{I} & 0\\
      \left(\bf J-2nn^{\rm T}J\right)  & 0 & 0\\
      \bf L & 0 & \left(\bf J- 2nn^{\rm T}J - n^{\rm T}Jn\mathbb{I}\right) 
    \end{array}\right]
\end{equation}
This means, eigenvalues of $\tilde{\bf J}_{1}$ can be obtained from the
eigenvalues of $\bf N = \left(\bf J-n^{T}Jn\mathbb{I} - 2nn^{\rm T}J\right)$ 
and square root of the eigenvalues of $\bf M = \left(\bf J - 2nn^{\rm T}J\right)$. 

Let $\lambda_{1}\left(\bf x\right)\le \lambda_{2}\left(\bf
  x\right)\le\lambda_{3}\left(\bf x\right)\le...\le\lambda_{n}\left(\bf
  x\right)$ be the eigenvalues (including degeneracies) of $\bf
H\left(\bf x\right)$ and $\left({\bf w}_{1},{\bf w}_{2},...,{\bf w}_{n}\right)$ be
the associated set of orthonormal eigenvectors at some fixed point of
$V$. A negative (positive) eigenvalue of $\bf J$ (Hessian) corresponds
to a stable manifold, while a positive eigenvalue of $\bf J$ corresponds to an
unstable manifold. Assuming $\bf n$ equals ${\bf w}_{1}$ (i.e. the
minimum eigen mode of the local Hessian), we find that
\begin{equation}
{\bf Mw}_{i} = \left(\bf J - 2nn^{\rm T}J\right){\bf w}_{i} 
= -\lambda_{i}{\bf w}_{i} + 2\lambda_{1}\delta_{i1}{\bf w}_{i}
\label{Dynamical5}
\end{equation}
\begin{equation}
{\bf Nw}_{i} = \left(\bf J-n^{T}Jn\mathbb{I} - 2nn^{\rm T}J\right){\bf w}_{i} 
= -\lambda_{i}{\bf w}_{i} + \lambda_{1}{\bf w}_{i} + 2\lambda_{i}\delta_{i1}{\bf w}_{i}
\label{Dynamical6}
\end{equation}

Thus, the eigenvalues of $\tilde{\bf J}_{1}$ are
$\{\sqrt{\lambda_{1}}, -\sqrt{\lambda_{1}}, 2\lambda_{1}\}$ for $i=1$ and $\{\sqrt{-\lambda_{i}},
-\sqrt{-\lambda_{i}}, \left(\lambda_{1}-\lambda_{i}\right)\}$ 
when ${\bf w}_{i}\perp {\bf w}_{1}$ (i.e. $i>1$). At an index-1
saddle point $\lambda_{1}<0$ and $\lambda_{i}>0$ for all $i>1$. Consequently, the eigenvalues of
$\tilde{\bf J}_{1}$ are all negative or complex numbers and this fixed
point of $V\left(\bf x\right)$ becomes linearly stable attractor for
the dynamical system ($\ref{EqMotion1}$). 

At a locally stable fixed point of $V\left(\bf x\right)$ all
the eigenvalues of $\bf H$ are positive. Thus, $\tilde{\bf J}_{1}$ has two
positive eigenvalues, $n$ negative eigenvalues and ($2n-2$) complex
eigenvalues. Correspondingly, there exists two unstable 
manifolds due to which this fixed point becomes unstable in MD-GAD.
Detailed analysis of convergence of GAD to index-1 saddle points for
non-gradient systems has been reported elsewhere.$\cite{EX11}$

Now, for the dynamical system ($\ref{EqMotion1}$) the Jacobian
near a fixed point of $V\left(\bf x\right)$ is given by 
\begin{equation}
  \tilde{\bf J}_{2} = \left[\begin{array}{cc}
      \left(\bf J-2nn^{\rm T}J\right) & 0\\
      {\bf L} & \left(\bf J- {2}nn^{\rm T}J - n^{\rm T}Jn\mathbb{I}\right)\\
    \end{array}\right]
  \label{Dynamical7}
\end{equation}
Following a similar procedure, the eigenvalues of the $\tilde{\bf
  J}_{2}$ are given by the eigenvalues of $\bf N$ and $\bf M$. We find
that all the eigenvalue of $\tilde{\bf J}_{2}$ are negative at 
index-1 saddle point of $V\left(\bf x\right)$. In contrast, there exists two positive
and (2$n$-2) negative eigenvalues of $\tilde{\bf J}_{2}$ at a stable fixed
point of $V\left(\bf x\right)$. Thus local minimum points of $V\left(\bf x\right)$ are linearly unstable
while index-1 saddle points become locally stable attractors. 

\section{Appendix II:Local convergence analysis for ART}
Having analyzed the stability of different fixed points of $V\left(\bf
  x\right)$ within GAD, let us try to analyze the stability of fixed 
points on $V\left(\bf x\right)$ in ART. Originally, ART was proposed to explore the PES
and only in subsequent modified versions it converges to a saddle
point.$\cite{CancesLMMW09, MarinicaWM11}$ Here we evaluate the
propensity of ART to visit an index-1 saddle point while exploring the PES.

In ART, the system evolves by the following equation:
\begin{equation}
  \bf \dot{x} = \bf F\left(\bf x\right) - \nu\left({\bf F}\left(\bf
      x\right),\bf n\right)n, \qquad \nu> 1\\
  \label{ART1}
\end{equation}
where, $\bf n$ is the direction vector given by the difference of the
current location $\bf x$ and initial location $\bf x_{o}$ in the configuration space,
i.e. $\bf n = \left(\bf x - x_{o}\right)/\left|\bf x -
  x_{o}\right|$. Close to a fixed point of $V\left(\bf x\right)$ since the force is
zero, the Jacobian, $\tilde{\bf J}_{3}$, for ($\ref{ART1}$) is
$\tilde{\bf J}_{3} = \left(\bf J-\nu nn^{\rm 
 T}J\right)$ where, ${\bf J} = -{\bf H} = -\nabla^{2} V\left(\bf x\right)$. To
illustrate the stability of the dynamical system in 
($\ref{ART1}$), let us consider for example a simple 2D potential 
energy landscape : 
\begin{equation}
  V\left(x,y\right) = \sin\left(\pi x\right)\sin\left(\pi y\right).
  \label{ART2}
\end{equation}
At a saddle point, for instance at ($0,0$), the Hessian has
eigenvalues $\{1, -1\}$. Assuming ${\bf n} =
\left[n_{1}\;\;n_{2}\right]^{\rm T}$, with $n^{2}_{1} + n^{2}_{2} =
1$, the Jacobian of ART at a fixed point of $V\left(\bf x\right)$ is  
\begin{equation}
  \tilde{\bf J}_{3} = \left(\left[\begin{array}{cc} 1 & 0\\ 0 &
      1\end{array}\right] - \nu \left[\begin{array}{cc} n^{2}_{1} &
      n_{1}n_{2}\\ n_{1}n_{2} & n^{2}_{2}\end{array}\right]\right)\bf J
  \label{ART3}
\end{equation}
At ($0,0$), if $\bf J$ is given by
\begin{equation}
  \bf J = \left[\begin{array}{cc} -1 & 0\\ 0 &
      1\end{array}\right],\quad\Rightarrow\; 
  \tilde{\bf J}_{3} = \left[\begin{array}{cc} \left(1-\nu
        n^{2}_{1}\right) & -\nu n_{1}n_{2}\\ \nu n_{1}n_{2} & 
      \left(1-\nu n^{2}_{2}\right)\end{array}\right]
  \label{ART4}
\end{equation}
Diagonalizing $\tilde{\bf J}_{3}$, we obtain the following eigenvalues 
\begin{equation}
  \left\{\frac{\left(2-\nu\right)}{2}-\frac{\nu}{2}\sqrt{\left(1-8n^{2}_{1}n^{2}_{2}\right)}\;,\;
  \frac{\left(2-\nu\right)}{2}+\frac{\nu}{2}\sqrt{\left(1-8n^{2}_{1}n^{2}_{2}\right)}\;\right\}
  \label{ART5}
\end{equation}
Since, $1 = \left(n^{2}_{1}+n^{2}_{2}\right)^{2}\ge
4n^{2}_{1}n^{2}_{2}\;\Rightarrow n^{2}_{1}n^{2}_{2}\le 1/4$,
consequently $\left(1-8n^{2}_{1}n^{2}_{2}\right)\ge
-1$. Thus, depending on the direction vector $\bf n$ the saddle can be
stable or unstable for the dynamical equation in ($\ref{ART1}$).

% \section{bibliography}
% \clearpage

\clearpage

\end{document}